\documentclass{aa}  
\usepackage[colorlinks=true, linkcolor=red, urlcolor=cyan, citecolor=blue]{hyperref}
\usepackage{graphicx}
\usepackage{txfonts}
\usepackage{multirow}
\usepackage{pdflscape}
\usepackage{rotate}

\usepackage{color}
\begin{document}

   \title{Slowly, slowly in the wind}
   \subtitle{3D hydrodynamical simulations of wind mass transfer and angular-momentum loss in AGB binary systems}

   \author{M. I. Saladino
          \inst{1, 2}
          \and
          O. R. Pols\inst{1}
          \and C. Abate\inst{3}
          }

\institute{Department of Astrophysics/IMAPP, Radboud University, P.O. Box 9010, 6500 GL Nijmegen, The Netherlands\\
              \email{[m.saladino; o.pols]@astro.ru.nl}
               \and
               Leiden Observatory, Leiden University, PO Box 9513, 2300, RA, Leiden, The Netherlands        
                \and
            	Argelander-Institute f\"{u}r Astronomie (AIfA), University of Bonn, Auf dem H\"{u}gel 71, 53121 Bonn, Germany
             \\
             	\email{cabate@uni-bonn.de}
             }

   \date{Received September 15, 2020; accepted March 16, 2020}

 
  \abstract
  {Wind mass transfer in binary systems with asymptotic giant branch (AGB) donor stars plays a fundamental role in the formation of a variety of objects, including barium stars and carbon-enhanced metal-poor (CEMP) stars. 
  In an attempt to better understand the properties of these systems, we carry out a comprehensive set of smoothed-particle hydrodynamics (SPH) simulations of wind-losing AGB stars in binaries, for a variety of binary mass ratios, orbital separations, initial wind velocities and rotation rates of the donor star. The initial parameters of the simulated systems are chosen to match the expected progenitors of CEMP stars. 
   We find that the strength of interaction between the wind and the stars depends on both the wind-velocity-to-orbital-velocity ratio ($v_\infty/v_\mathrm{orb}$) and the binary mass ratio. 
   Strong interaction occurs for close systems and comparable mass ratios, and gives rise to a complex morphology of the outflow and substantial angular-momentum loss, which leads to a shrinking of the orbit. 
   As the orbital separation increases and the mass of the companion star decreases, the morphology of the outflow, as well as the angular-momentum loss, become more similar to the spherically symmetric wind case. 
   We also explore the effects of tidal interaction and find that for orbital separations up to 7-10 AU, depending on mass ratio, spin-orbit coupling of the donor star occurs at some point during the AGB phase. If the initial wind velocity is relatively low, we find that corotation of the donor star results in a modified outflow morphology that resembles wind Roche-lobe overflow.
   In this case the mass-accretion efficiency and angular-momentum loss differ from those found for a non-rotating donor. 
   Finally, we provide a relation for both the mass-accretion efficiency and angular-momentum loss as a function of $v_\infty/v_\mathrm{orb}$ and the binary mass ratio that can be easily implemented in a population synthesis code to study populations of barium stars, CEMP stars and other products of interaction in AGB binaries, such as cataclysmic binaries and supernovae type Ia. 
	}

   \maketitle
%

\section{Introduction} \label{p2:sec:introd}

In recent years, major efforts have been made to characterise the properties of the outflow of nearby asymptotic giant branch (AGB) stars \citep[e.g.][]{decin+2010, blind+2011, jeffers+2014, decin+2015, kervella+2016}.
These studies shed unprecedented light on the structure of the winds, their chemical composition and dust-to-gas ratios, wind velocity profiles and mass-loss rates.
In many cases, they reveal the presence of a binary companion which interacts with the outflow and modifies its structure \citep{blind+2011, jeffers+2014, decin+2015, Bujarrabal+2018}. 
As many AGB stars are known to be in binary systems, several authors have attempted to model the complex interactions between the outflow and the two stellar components, with the aim of reproducing their observed morphologies \citep[e.g.][]{shazrene1, Kim+2017, Ramstedt+2017} and to understand how the evolution of the system is affected. 
To investigate the latter, several authors have studied what fraction of wind material can be transferred onto the low-mass companion \citep{theuns2, Nagae+2004,  val-borro, shazrene1, shazrene, Liu+2017, Chen+2017, Val-Borro+2017} and how much angular momentum is lost by wind ejection \citep{tavani, jahanara2, rochester, Saladino+2018a}.

A thorough understanding of wind mass transfer in binary systems is of fundamental importance to determine the formation mechanism of a variety of objects, such as symbiotic binaries \citep{Merrill1942, Merrill1944, Merrill1948, Merrill1950, Kenyon1992}, barium (Ba) stars \citep{Bidelman+Keenan1951}, carbon and s-process enhanced metal-poor (CEMP-$s$) stars \citep{Beers+Christlieb2005}, and blue stragglers in old open clusters \citep{mathieu}.
Binary population-synthesis studies typically assume that the AGB star loses mass in the form of a fast spherically symmetric wind and that the mass accretion process is described by the Bondi-Hoyle-Lyttlelton (BHL) formalism \citep{Hoyle+Lyttleton1939, Bondi+Hoyle1944, edgar}.
In this simplified model, the orbit of the binary expands in response to mass loss. 
As a consequence of these assumptions, theoretical studies of the progeny of AGB binary systems, such as CEMP-$s$ and Ba stars, predict a dearth of systems with orbital periods between a few hundred and a few thousand days \citep{pols, Izzard+2009, izzard, nie, Abate+2018}
which is at odds with the observations \citep[e.g.][]{jorissen+1998, jorissen2, Hansen+2016}.
In addition, in order to reproduce the chemical abundances of some of the progeny of AGB binary systems, mass-accretion efficiencies much larger than those predicted by the BHL formalism are needed \citep[e.g. see ][for CEMP stars]{carlo2, carlo4}.

In \citet[hereinafter Paper I][]{Saladino+2018a}, we performed smoothed particle hydrodynamic (SPH) simulations including cooling of the gas to model interactions of non-rotating AGB binary systems via their winds. 
We used a predefined constant wind velocity and adopted the same initial stellar parameters of the binary components in all models. 
We studied the effect of the terminal velocity of the wind on the mass-accretion efficiency and on the angular-momentum loss.
We found that these quantities show a strong dependence on the ratio of the wind velocity to the relative orbital velocity, $v_{\infty}/v_\mathrm{orb}$.
Despite differences in the assumptions, by comparing to the hydrodynamical models by \cite{jahanara2} and \cite{Chen+2017}, we found a similar trend for the angular-momentum loss as a function of $v_{\infty}/v_\mathrm{orb}$. 
In Paper I, we also presented a preliminary analytical expression which describes the angular-momentum loss in terms of $v_{\infty}/v_\mathrm{orb}$.

In the context of CEMP-$s$ stars, \cite{Abate+2018} performed binary population synthesis simulations in an attempt to reproduce the observed orbital-period distribution of the CEMP stars in the sample obtained by \cite{Hansen+2016}.
In the simulations of \cite{Abate+2018} mass-accretion efficiency and angular-momentum are treated independently, because hydrodynamical simulations computing simultaneously the mass transfer rates and the angular-momentum loss via winds were not available at the time. 
In their model sets M6 and M7, \cite{Abate+2018} implement the fit we provide in Paper I for angular-momentum loss, and adopt the wind Roche-lobe overflow (WRLOF) model for the mass-accretion efficiency \citep{shazrene1, carlo1}. 
Although with these assumptions they find a slight increase in the proportion of evolved binaries at short orbital periods (< 2500 day) compared to their default model set assuming isotropic winds, the period distribution of their synthetic population does not reproduce the observations. 
A self-consistent model in which angular-momentum loss and mass-accretion efficiencies are calculated from simulations where the same set of conditions are adopted may help to improve the results of binary population synthesis codes on the populations of CEMP-$s$ stars. 
In addition, most hydrodynamical studies have focused on investigating the consequences of modifying one orbital parameter at a time on the mass-accretion efficiency or angular-momentum loss. 
No hydrodynamical study has yet been performed with the same set of assumptions for a large grid of binary and wind parameters. 
The aim of this work is to perform such hydrodynamical simulations for a much wider range of binary and wind parameters, in order to have a better understanding of the important quantities driving the interaction of AGB winds with the binary systems.

Another aspect that may be of importance and that is sometimes overlooked is the effect of the spin of the donor star on the interacting binary.
During the AGB phase the rotational velocity of a single star is greatly reduced due to the expansion of its envelope and because angular momentum is carried away by the mass the star is losing. 
For this reason, the angular velocity of a single AGB star is negligible small.
However, when the AGB star is in a binary system tidal effects can trigger spin-orbit coupling. 
In the equilibrium-tide model the synchronisation timescale depends on the ratio $(R_\mathrm{d}/a)^{-6}$, where $R_\mathrm{d}$ is the radius of the donor star and $a$ the orbital separation of the binary \citep{zahn1977}. 
This implies that for AGB binaries in which the donor star is close to filling its Roche lobe, tidal effects can be important since the synchronization timescale is usually much shorter than the typical evolution timescale of an AGB star. 
Most of the previous hydrodynamical studies of AGB binaries are performed in the corotating frame, thus corotation of the AGB star is implicit in the simulations. 
Only a few works \citep[e.g.][]{theuns2, Saladino+2018a} have performed hydrodynamical simulations in the inertial frame of the binary system, assuming the donor star is not rotating.
The morphologies found in some of the models in which corotation of the donor is assumed, correspond to the WRLOF geometry, i.e. the dense wind is confined to the Roche lobe of the donor star and flows through the inner Lagrange point towards the companion star \citep[e.g.][]{val-borro, shazrene, Chen+2017}.
In contrast, the morphologies found when considering a non-rotating donor star are somewhat different, mainly showing spiral patterns formed by the accretion wake of the companion star. 
The models by \cite{shazrene} and \cite{Chen+2017} show that when interaction of the binary stars occurs via WRLOF the mass-accretion efficiency onto the companion star is enhanced.
Considering the differences in the results found in numerical models, it is necessary to better understand how rotation modifies the morphology of the AGB outflow in binary systems, and to constrain the range of orbital parameters within which it is realistic to assume corotation of the AGB donor. 

In this work we present a comprehensive study of how angular-momentum loss and accretion efficiency in low-mass binary stars vary as a function of mass ratio, orbital separation, and wind-to-orbital velocity ratios. 
We provide two analytical expressions for angular-momentum loss and mass-accretion efficiency that can be easily implemented in binary population synthesis codes .
In addition, we explore how the results are changed if we assume that the AGB donor is corotating with the binary system instead of non-rotating and under which circumstances corotation can be expected. 

\section{Method}\label{p2:sec:method}

\subsection{Numerical method}\label{p2:sec:numerical_method}

The numerical method used in this paper for the SPH simulations is similar to that described in Paper I. 
Here we will only give a brief summary.
We use the Astrophysical Multi-purpose Software Environment \textsc{amuse} \citep{amuse3,amuse2,amuse1} to couple the smoothed particle hydrodynamics (SPH) code \textsc{fi} \citep{fi1,fi2,fi3} with the N-body code \textsc{huayno} \citep{huayno} using the \textsc{Bridge} module \citep{bridge}. We use the N-body code to evolve the orbits of the stars. 
We choose an SPH code because these codes are arguably better at conserving angular momentum than grid-based codes \citep{price_mother}.
Unlike in Paper I, we allow the stars to feel the gravitational field of the gas and the self-gravity of the gas is taken into account. 

The AGB wind is modelled using the "accelerating wind" mode of the \textsc{stellar\_wind.py} routine \citep[][also available in \textsc{amuse}]{edwin}.  
Wind particles are created in a shell around the star losing mass and are injected with initial velocity $v_\mathrm{init}$. The initial temperature of the wind particles is equal to the effective temperature of the donor star $T_\mathrm{eff}$. The acceleration given to the SPH particles balances the gravity of the donor star: 
\begin{equation}
\label{p2:eq:acceleration}
a_\mathrm{w}(r) = \frac{GM_\mathrm{d}}{r^2}.
\end{equation}
Due to gas pressure, the wind will experience an outward acceleration close to the donor star until it reaches its terminal velocity $v_{\infty}>v_\mathrm{init}$ at some distance from the surface of the AGB star.
This acceleration mechanism differs from Paper I, in which the acceleration was set to compensate for each term in the equation of motion of the gas, such that the velocity of the wind was constant as a function of distance form the star ($dv/dr=0$).  

Similar to Paper I, the gas is assumed to be monoatomic, with an adiabatic index $\gamma = 5/3$. The equation governing the cooling or heating of the gas is given by:
\begin{equation}
\label{p2:eq:bowen_cooling}
\dot{Q} = \frac{3k_\mathrm{B}}{2\mu m_{u}} \frac{(T-T_\mathrm{eq})\rho}{C} + \dot{Q}_\mathrm{rad}, 
\end{equation}
where $k_\mathrm{B}$ is the Boltzmann constant, $T$ is the gas temperature, $\rho$ is the gas density, $\mu$  is the mean molecular weight and $C$ a constant parameter which value is $10^{-5}$ g s cm$^{-3}$. The the first term assumes that cooling comes from gas radiating away or absorbing thermal energy trying to reach the equilibrium temperature at radius $r$ given by the Eddington approximation for a gray spherical atmosphere \citep{chandrasekhar1}:
\begin{equation}
\label{p2:eq:grey_atmosphere}
T^4_\mathrm{eq} = \frac{1}{2} T_\mathrm{eff}^{4} \left\{ \left[1-\left(1-\frac{R_\mathrm{d}^2}{r^2}\right)^{1/2}\right] + \frac{3}{2} \int_{r}^{\infty} \frac{R_\mathrm{d}^2}{r^2}(\kappa_\mathrm{g} + \kappa_\mathrm{d}) \rho dr\right\}, 
\end{equation}
where  $\kappa_\mathrm{g}$ is the gas opacity and $\kappa_\mathrm{d}$ is the dust opacity.
Unlike Paper I, where we assumed a constant value of the total opacity ($\kappa_\mathrm{g}$ + $\kappa_\mathrm{d}$), in this work we only allow for the dust opacity to come into play at distances larger than three times the radius of the donor star (3$R_\mathrm{d}$), where dust is expected to form according to AGB wind models \citep{Hoefner+2018}. 
The values of the opacities are constant in the calculations and equalt to $\kappa_\mathrm{g} = 2 \times 10^{-4}$ cm$^{2}$ g$^{-1}$ and $\kappa_\mathrm{d} = 5$ cm$^{2}$ g$^{-1}$ \citep{bowen}. 
The second term in Equation \ref{p2:eq:bowen_cooling} corresponds to the cooling rate prescription for high temperatures ($\log T \geq 3.8$ K) of \citet{schure}. 
Because the abundances in this paper are different from those in Paper I (see section \ref{p2:sec:binary_setup}), we compute a new cooling table using equation 3 of \citet{schure} and the abundances $n_i$ from Table \ref{p2:table:abundances}.

The resolution in SPH is defined by the smoothing length $h_{i} \propto (m_\mathrm{g}/\rho_i)^{1/3}$, where $m_\mathrm{g}$ is the SPH gas particle mass and $\rho_i$ the density of the gas at the position $\mathbf{r}_i$ of the particle $i$ . 
In Paper I, we found that the angular-momentum loss is independent of the resolution used, while the accretion efficiency is only weakly dependent on the resolution (see Figure 2 of that paper).
Because of this, and in order to minimize the computational time, the resolution we use in this work corresponds to that used in model R3 of Paper I. 
For an orbital separation of 5 AU, and scaling the density to the value of the mass-loss rate, $\dot{M}_\mathrm{d}$, used (see Section \ref{p2:sec:binary_setup}), the corresponding mass of an SPH particle is $m_\mathrm{g} = 1.2 \times 10^{-9}$ M$_{\odot}$. 
Furthermore, in order to optimize the numerical computation at large orbital separations, we choose the typical smoothing length to be proportional to the semi-major axis of the binary, $a$. 
Since the average gas density is expected to decrease with the inverse square of the distance to the donor star, i. e. to scale with $a^{-2}$, we can achieve this by choosing the SPH particle mass $m_\mathrm{g}$ to be proportional to $a$. 

As seen in Paper I (see Figure 1 of that paper) the transfer of angular momentum from the binary orbit to the gas occurs within a few times the orbital separation. 
For this reason, and in order to minimize the computational time, we remove the SPH particles once they have crossed a boundary at $5a$ from the centre of mass of the system.
Finally, similar to Paper I the values of the artificial viscosity parameters are $\alpha_\mathrm{SPH} = 0.5$ and $\beta_\mathrm{SPH} = 1$. As shown in Paper I, adopting different values for these parameters does not modify the results in a significant way.

\subsection{Binary set up}\label{p2:sec:binary_setup}

The motivation of this paper is to study the effect of angular-momentum loss and mass-accretion efficiency for a range of parameters relevant for low-mass binary stars. In this work we explore the low-metallicity case of the donor star, which applies to the progenitors of CEMP-$s$ stars. 
The low metallicity will impact several properties of the donor star during the AGB phase, such as stellar radius, effective temperature and consequently the mass-loss rate. 
In order to obtain realistic parameters for the AGB donor, we compute the evolution of a single star of initial mass $M_\mathrm{d, init} = 1.5$ M$_{\odot}$ and initial metallicity $Z_\mathrm{init} = 10^{-4}$ using the population nucleosynthesis code \texttt{binary\_c}\footnote{SVN revision No. 5045} \citep{Izzard+2004, Izzard+2006, Izzard+2009, carlo1, Izzard+2018}. The initial metal abundances of the star are equal to solar \citep{asplund} scaled down to $Z_\mathrm{init}$. 
The mass-loss rate of the AGB star is calculated using the method described in \citet{VassiliadisWoods1993}.
As starting conditions for the simulations, we adopt the stellar mass, $M_\mathrm{d}$, radius, $R_\mathrm{d}$,
effective temperature, $T_\mathrm{eff}$, mass-loss rate, $\dot{M}_\mathrm{d}$, and chemical abundances at the moment the star has reached the superwind phase
on the AGB, and the radius is close to its maximum value. 
This is the phase of evolution when the strongest interaction between the wind and the companion will occur.
At this moment the mass is reduced to 1.2 M$_{\odot}$, its age is 1796 Myr and its surface is strongly enriched in carbon (see Tables \ref{p2:table:donor_star} and \ref{p2:table:abundances}). 

\begin{table}[h]
\centering
\caption{Constant parameters of the donor star in the hydrodynamical simulations}
\label{p2:table:donor_star}
\begin{tabular}{l l l}
\hline\hline
Parameter	&	Value	&	Description	\\ \hline
$M_\mathrm{d}$	&	1.2 M$_{\odot}$	&	Mass of the AGB star	\\
$R_\mathrm{d}$	&	330 R$_{\odot}$	&	Radius of the AGB star,\\
 & & and inner boundary for\\
 & & release of new SPH particles\\
$\dot{M}_\mathrm{d}$	&	$1.5 \times 10^{-5}$ M$_{\odot}$ yr$^{-1}$	&	mass-loss rate 	\\
$T_\mathrm{eff}$	&	3240	&	Effective temperature	\\ \hline \hline
\end{tabular}
\end{table}

Similar to Paper I, the companion star is modelled as a sink particle with radius equal to  $0.1 R_\mathrm{L,2}$, where $R_\mathrm{L,2}$ is the Roche-lobe radius of the accretor star. 
For the companion star mass, $M_\mathrm{a}$, we take $M_\mathrm{a} = 1.2$ M$_{\odot}$, 0.9 M$_{\odot}$, 0.6 M$_{\odot}$, 0.4 M$_{\odot}$, and 0.3 M$_{\odot}$.
The mass ratio, $q = M_\mathrm{d}/M_\mathrm{a}$, varies accordingly between 1 and 4.
These mass ratios encompass the mass ratios of CEMP-$s$ progenitors from \citet{carlo2, carlo3}.
We should note that these are not the initial mass ratios, but the mass ratios at the time when AGB mass transfer takes place. 

Using the set up described above, we perform 31 simulations (Table \ref{p2:table:setup}) in which the initial orbit of the binary system is circular and its initial orbital  separation ranges between 4 and 20 AU. 
These orbital separations are chosen so that they correspond to the typical orbital periods of CEMP-$s$ progenitor stars found by \citet{carlo2, carlo3}. 
The stars are placed in the inertial frame and the centre of mass of the binary system is positioned at the origin. 
We simulate models with different initial velocities of the wind $v_\mathrm{init} = 1$, 5 and 12 km s$^{-1}$. 
We note that in the models with the closest orbits (4-5 AU) the donor star fills a substantial fraction of its Roche lobe and its surface layers and atmosphere will be deformed by tides from the companion to some extent. 
We do not take this tidal deformation into account in the simulations, since the wind particles are always injected from a spherical shell with inner radius $R_d$ (see Section \ref{p2:sec:numerical_method}).

\subsection{Rotation and spin-orbit coupling} \label{p2:sec:method_rot}

In this paper we explore the possibility of rotation of the AGB donor.
In order to check for which orbital separations corotation of the AGB donor with the orbit is expected, we perform a set of simulations within \texttt{binary\_c} of binary systems interacting via isotropic winds and we compute their tidal evolution. 
The physics of tidal evolution in \texttt{binary\_c} is implemented following the work of \citet[Section 2.3 and references therein]{hurley} and the equation for the tidal circularisation and synchronisation timescale is based on \cite{Rasio+1996}. 
The equations of tidal evolution are originally reported by \cite{Hut1981}. 
We evolve systems with initial orbital separations between 3 and 30 AU, initial masses as in Section \ref{p2:sec:binary_setup} and with the initial angular velocity of the donor star, $\Omega_\mathrm{spin}$, set equal to zero, i.e. the star is initially non-rotating. 
We check the angular velocity of the donor star at the time the mass of the AGB star and the mass ratios are equal to those assumed in our hydrodynamical simulations. 
We find that for orbital separations up to $\approx$ 15-20 AU, depending on the mass ratio, tidal interactions have spun up the donor star to a non-zero angular velocity.
For initial separations smaller than 7-10 AU a state of near-corotation, with $\Omega_\mathrm{spin}$ comparable to the angular velocity of the binary, $\Omega_\mathrm{bin}$, is reached at some point during the AGB phase. 
However, except for very close orbital separations ($a \la 4-6$ AU), corotation is lost again by the time we start our hydrodynamical simulations as a result of the high mass-loss rate of the AGB star. 
The ratio $\Omega_\mathrm{spin}/\Omega_\mathrm{bin}$ at this moment is a decreasing function of separation and mass ratio (see Appendix \ref{p2:appendix:rot} for details).

So as to account for spin-orbit coupling, and in order to investigate the effect of rotation on the angular-momentum loss and mass-accretion efficiency, in a subset of our simulations we assume both non-rotation and corotation with the orbit for models with the same mass ratio and the same orbital separation (see Table \ref{p2:table:setup}; the corotating models are designated by the symbol "$\Omega$"). 
In the models where corotation of the binary is assumed, we add a tangential component to the velocity of the wind particles as they leave the donor star,  $\mathbf{v}_\mathrm{T} = \mathbf{\Omega}_{\mathrm{spin}} \times \mathbf{r}$, with $\Omega_\mathrm{spin}=\Omega_\mathrm{bin}$ and $\mathbf{r}$ the position of the gas particles with respect to the centre of mass of the donor star. 
For the other simulations we assume the donor star to be non-rotating regardless the fact that $\Omega_\mathrm{spin}$ may be non-zero.

\begin{table*}[h]
\centering
\caption{Parameters of the simulations}
\label{p2:table:setup}
\begin{tabular}{c c c c c c c c c c c c}
\hline\hline																						
Model		&	$M_\mathrm{a}$	&	$a$		&	$P_\mathrm{orb}$		&	$q$		&	 $\Omega_\mathrm{spin}$																				&	$v_\mathrm{init}$		&	$v_\mathrm{orb}$ 	&	$R_\mathrm{L,1}$	&	$R_\mathrm{L,2}$	&	$m_\mathrm{g}$		\\ 
-	&	M$_{\odot}$	&	AU	&	yr	&	-	&	s$^{-1}$	&	km s$^{-1}$	&	km s$^{-1}$	&	AU	&	AU	&	M$_{\odot}$	\\ \hline
Q1a5v1	&	1.2	&	5	&	7.22	&	1	&	0	&	1	&	20.64	&	1.92	&	1.92	&	$1.2\times10^{-9}$	\\
Q1a5v5	&	1.2	&	5	&	7.22	&	1	&	0	&	5	&	20.64	&	1.92	&	1.92	&	$1.2\times10^{-9}$	\\
Q1a5$\Omega$	&	1.2	&	5	&	7.22	&	1	&	$2.76 \times 10^{-8}$	&	12	&	20.64	&	1.92	&	1.92	&	$1.2\times10^{-9}$	\\
Q1a5	&	1.2	&	5	&	7.22	&	1	&	0	&	12	&	20.64	&	1.92	&	1.92	&	$1.2\times10^{-9}$	\\
Q1a10	&	1.2	&	10	&	20.41	&	1	&	0	&	12	&	14.59	&	3.83	&	3.83	&	$2.4\times10^{-9}$	\\
Q1a20	&	1.2	&	20	&	57.73	&	1	&	0	&	12	&	10.32	&	7.66	&	7.66	&	$4.8\times10^{-9}$	\\
Q13a5v1	&	0.9	&	5	&	7.71	&	4/3	&	0	&	1	&	19.30	&	2.04	&	1.79	&	$1.2\times10^{-9}$	\\
Q13a5v5	&	0.9	&	5	&	7.71	&	4/3	&	0	&	5	&	19.31	&	2.04	&	1.79	&	$1.2\times10^{-9}$	\\
Q13a5	&	0.9	&	5	&	7.71	&	4/3	&	0	&	12	&	19.30	&	2.04	&	1.79	&	$1.2\times10^{-9}$	\\
Q13a10	&	0.9	&	10	&	21.82	&	4/3	&	0	&	12	&	13.65	&	4.08	&	3.58	&	$2.4\times10^{-9}$	\\
Q13a20	&	0.9	&	20	&	61.73	&	4/3	&	0	&	12	&	9.65	&	8.17	&	7.16	&	$4.8\times10^{-9}$	\\
Q2a4v1$\Omega$	&	0.6	&	4	&	5.96	&	2	&	$3.34 \times 10^{-8}$	&	1	&	19.98	&	1.78	&	1.29	&	$9.6\times10^{-10}$	\\
Q2a4v1	&	0.6	&	4	&	5.96	&	2	&	0	&	1	&	19.98	&	1.78	&	1.29	&	$9.6\times10^{-10}$	\\
Q2a5v5$\Omega$	&	0.6	&	5	&	8.33	&	2	&	$2.39 \times 10^{-8}$	&	5	&	17.87	&	1.78	&	1.29	&	$1.2\times10^{-9}$	\\
Q2a5v5	&	0.6	&	5	&	8.33	&	2	&	0	&	5	&	17.87	&	1.78	&	1.29	&	$1.2\times10^{-9}$	\\
Q2a5$\Omega$	&	0.6	&	5	&	8.33	&	2	&	$2.39 \times 10^{-8}$	&	12	&	17.87	&	2.22	&	1.61	&	$1.2\times10^{-9}$	\\
Q2a5	&	0.6	&	5	&	8.33	&	2	&	0	&	12	&	17.87	&	2.22	&	1.61	&	$1.2\times10^{-9}$	\\
Q2a10	&	0.6	&	10	&	23.57	&	2	&	0	&	12	&	12.64	&	4.44	&	3.23	&	$2.4\times10^{-9}$	\\
Q2a20	&	0.6	&	20	&	66.66	&	2	&	0	&	12	&	8.94	&	8.88	&	6.46	&	$4.8\times10^{-9}$	\\
Q3a4v1	&	0.4	&	4	&	6.32	&	3	&	0	&	1	&	18.84	&	1.92	&	1.16	&	$9.6\times10^{-10}$	\\
Q3a5v5	&	0.4	&	5	&	8.83	&	3	&	0	&	5	&	16.85	&	2.40	&	1.45	&	$1.2\times10^{-9}$	\\
Q3a4	&	0.4	&	4	&	6.32	&	3	&	0	&	12	&	18.84	&	1.92	&	1.16	&	$9.6\times10^{-10}$	\\
Q3a5	&	0.4	&	5	&	8.83	&	3	&	0	&	12	&	16.85	&	2.40	&	1.45	&	$1.2\times10^{-9}$	\\
Q3a10	&	0.4	&	10	&	25.00	&	3	&	0	&	12	&	11.91	&	4.79	&	2.89	&	$2.4\times10^{-9}$	\\
Q3a20	&	0.4	&	20	&	70.7	&	3	&	0	&	12	&	8.42	&	9.58	&	5.78	&	$4.8\times10^{-9}$	\\
Q4a4v1	&	0.3	&	4	&	6.53	&	4	&	0	&	1	&	18.24	&	2.52	&	1.33	&	$1.2\times10^{-9}$	\\
Q4a5v5	&	0.3	&	5	&	9.13	&	4	&	0	&	5	&	16.32	&	2.52	&	1.33	&	$1.2\times10^{-9}$	\\
Q4a4	&	0.3	&	4	&	6.53	&	4	&	0	&	12	&	18.24	&	2.02	&	1.07	&	$9.6\times10^{-10}$	\\
Q4a5$\Omega$	&	0.3	&	5	&	9.13	&	4	&	$2.18 \times 10^{-8}$	&	12	&	16.32	&	2.52	&	1.33	&	$1.2\times10^{-9}$	\\
Q4a5	&	0.3	&	5	&	9.13	&	4	&	0	&	12	&	16.32	&	2.52	&	1.33	&	$1.2\times10^{-9}$	\\
Q4a10	&	0.3	&	10	&	25.82	&	4	&	0	&	12	&	11.54	&	5.04	&	2.66	&	$2.4\times10^{-9}$	\\																							
\hline \hline
\end{tabular}
\tablefoot{
$M_\mathrm{a}$ is the mass of accretor star.
$a$ is the orbital separation of the binary.
$P_\mathrm{orb}$	 the orbital period.
$q$ is the mass ratio $M_\mathrm{d}/M_\mathrm{a}$.
$\Omega_\mathrm{spin}$ is the angular velocity added to the gas particles to mimic the rotation of the donor star. It is the same as the angular velocity of the binary.
$v_\mathrm{init}$	 is the initial velocity with which particles are injected at the radius of the donor star.
$v_\mathrm{orb}$ is the relative orbital velocity of the binary system.
$R_\mathrm{L,1}$ is the size of the Roche-lobe of the donor star.
$R_\mathrm{L,2}$ is the size of the Roche-lobe of the companion star.
$m_\mathrm{g}$	is the mass of the SPH gas particles.
}
\end{table*}

\section{Results}\label{p2:sec:results}

\begin{table*}[h]
\centering
\caption{Results from the simulations}
\label{p2:table:results}
\begin{tabular}{c c c c c c c c c c}
\hline\hline
Model		&	$v_\infty$	&	$v_\mathrm{RL,1}$	&	$v_\infty/v_\mathrm{orb}$	&	$\eta$	&	$\beta$	&	$\beta_\mathrm{BHL}$	&	$	(\dot{a}/a)_\mathrm{bin}	$	&	$	(\dot{a}/a)_\mathrm{dyn}	$	&	Accretion disk	\\ 
-	&	km s$^{-1}$	&	km s$^{-1}$	&	-	&	-	&	-	&	-	&		yr$^{-1}	$	&		yr$^{-1}	$	&	-	\\ \hline
Q1a5v1	&	6.0	&	2.1	&	0.291	&	0.610	&	0.298	&	0.571	&	$	-8.18	\times 10^{-6}$	&	$	-9.11	\times 10^{-6}$	&	Yes	\\
Q1a5v5	&	10.9	&	7.3	&	0.526	&	0.607	&	0.305	&	0.245	&	$	-8.08	\times 10^{-6}$	&	$	-7.43	\times 10^{-6}$	&	Yes	\\
Q1a5$\Omega$	&	15.1	&	12.5	&	0.733	&	0.557	&	0.212	&	0.135	&	$	-7.17	\times 10^{-6}$	&	$	-3.73	\times 10^{-6}$	&	Yes	\\
Q1a5	&	15.1	&	12.5	&	0.732	&	0.575	&	0.223	&	0.135	&	$	-7.79	\times 10^{-6}$	&	$	-4.86	\times 10^{-6}$	&	Yes	\\
Q1a10	&	15.1	&	14.1	&	1.035	&	0.376	&	0.099	&	0.061	&	$	-6.58	\times 10^{-8}$	&	$	9.91	\times 10^{-7}$	&	No	\\
Q1a20	&	15.1	&	14.5	&	1.463	&	0.308	&	0.033	&	0.023	&	$	3.26	\times 10^{-6}$	&	$	3.94	\times 10^{-6}$	&	No	\\
Q13a5v1	&	6.0	&	2.1	&	0.311	&	0.548	&	0.323	&	0.386	&	$	-1.23	\times 10^{-5}$	&	$	-1.12	\times 10^{-5}$	&	Yes	\\
Q13a5v5	&	10.9	&	7.8	&	0.562	&	0.584	&	0.263	&	0.161	&	$	-1.42	\times 10^{-5}$	&	$	-1.02	\times 10^{-5}$	&	Yes	\\
Q13a5	&	15.1	&	13.0	&	0.782	&	0.363	&	0.125	&	0.086	&	$	-3.93	\times 10^{-6}$	&	$	-7.03	\times 10^{-7}$	&	No	\\
Q13a10	&	15.1	&	14.2	&	1.106	&	0.275	&	0.059	&	0.038	&	$	1.23	\times 10^{-6}$	&	$	2.38	\times 10^{-6}$	&	No	\\
Q13a20	&	15.1	&	14.5	&	1.565	&	0.220	&	0.016	&	0.014	&	$	4.80	\times 10^{-6}$	&	$	5.24	\times 10^{-6}$	&	No	\\
Q2a4v1$\Omega$	&	6.0	&	2.3	&	0.300	&	0.474	&	0.119	&	0.244	&	$	-1.96	\times 10^{-5}$	&	$	-1.61	\times 10^{-5}$	&	Yes	\\
Q2a4v1	&	6.0	&	2.3	&	0.300	&	0.515	&	0.263	&	0.244	&	$	-2.32	\times 10^{-5}$	&	$	-1.45	\times 10^{-5}$	&	Yes	\\
Q2a5v5$\Omega$	&	10.9	&	6.0	&	0.607	&	0.328	&	0.107	&	0.107	&	$	-9.76	\times 10^{-6}$	&	$	-4.36	\times 10^{-6}$	&	Yes	\\
Q2a5v5	&	10.9	&	6.0	&	0.607	&	0.381	&	0.128	&	0.085	&	$	-1.36	\times 10^{-5}$	&	$	-5.61	\times 10^{-6}$	&	Yes	\\
Q2a5$\Omega$	&	15.1	&	12.9	&	0.845	&	0.219	&	0.073	&	0.044	&	$	-1.60	\times 10^{-6}$	&	$	3.35	\times 10^{-7}$	&	No	\\
Q2a5	&	15.1	&	12.9	&	0.845	&	0.214	&	0.067	&	0.044	&	$	-1.08	\times 10^{-6}$	&	$	1.91	\times 10^{-7}$	&	No	\\
Q2a10	&	15.1	&	14.2	&	1.195	&	0.163	&	0.025	&	0.018	&	$	3.74	\times 10^{-6}$	&	$	4.60	\times 10^{-6}$	&	No	\\
Q2a20	&	15.1	&	14.5	&	1.689	&	0.131	&	0.007	&	0.007	&	$	6.63	\times 10^{-6}$	&	$	6.95	\times 10^{-6}$	&	No	\\
Q3a4v1	&	6.0	&	2.1	&	0.318	&	0.297	&	0.154	&	0.127	&	$	-1.79	\times 10^{-5}$	&	$	-1.08	\times 10^{-5}$	&	No	\\
Q3a5v5	&	10.9	&	8.6	&	0.644	&	0.160	&	0.052	&	0.043	&	$	-2.96	\times 10^{-6}$	&	$	-2.35	\times 10^{-6}$	&	No	\\
Q3a4	&	15.1	&	12.5	&	0.801	&	0.138	&	0.040	&	0.028	&	$	-3.13	\times 10^{-7}$	&	$	9.46	\times 10^{-7}$	&	No	\\
Q3a5	&	15.1	&	13.2	&	0.896	&	0.127	&	0.031	&	0.022	&	$	1.32	\times 10^{-6}$	&	$	2.51	\times 10^{-6}$	&	No	\\
Q3a10	&	15.1	&	14.3	&	1.268	&	0.091	&	0.010	&	0.009	&	$	5.96	\times 10^{-6}$	&	$	6.58	\times 10^{-6}$	&	No	\\
Q3a20	&	15.1	&	14.5	&	1.793	&	0.073	&	0.003	&	0.003	&	$	8.17	\times 10^{-6}$	&	$	8.48	\times 10^{-6}$	&	No	\\
Q4a4v1	&	6.0	&	2.6	&	0.329	&	0.152	&	0.048	&	0.078	&	$	-6.54	\times 10^{-6}$	&	$	-4.47	\times 10^{-6}$	&	No	\\
Q4a5v5	&	10.9	&	8.7	&	0.665	&	0.110	&	0.032	&	0.026	&	$	-1.20	\times 10^{-6}$	&	$	-5.00	\times 10^{-7}$	&	No	\\
Q4a4	&	15.1	&	12.4	&	0.828	&	0.093	&	0.022	&	0.017	&	$	1.65	\times 10^{-6}$	&	$	2.87	\times 10^{-6}$	&	No	\\
Q4a5$\Omega$	&	15.1	&	13.4	&	0.925	&	0.084	&	0.019	&	0.013	&	$	3.02	\times 10^{-6}$	&	$	4.39	\times 10^{-6}$	&	No	\\
Q4a5	&	15.1	&	13.4	&	0.925	&	0.083	&	0.017	&	0.013	&	$	3.29	\times 10^{-6}$	&	$	4.37	\times 10^{-6}$	&	No	\\
Q4a10	&	15.1	&	14.4	&	1.308	&	0.058	&	0.005	&	0.005	&	$	7.27	\times 10^{-6}$	&	$	7.67	\times 10^{-6}$	&	No	\\
\hline\hline
\end{tabular}
\tablefoot{			
$v_\infty$	is the average velocity of the wind at a distance 25$R_{\mathrm{d}}$ from a single star. 
$v_\mathrm{RL,1}$ is the average velocity of the wind at the Roche lobe of the donor star. We should note that this average velocity is taken assuming a single star undergoing mass loss via winds. 
$v_\infty/v_\mathrm{orb}$	is the velocity ratio of the terminal velocity of the wind to the relative orbital velocity of the binary system.
$\eta$	is the average specific angular momentum-loss of the mass that crossed a boundary at $3a$ from the binary system in units of $a^2\Omega_\mathrm{bin}$ as derived from Equation \ref{p2:eq:eta}.
$\beta$ is the average mass-accretion efficiency measured from the inflow in a shell at $0.4 R_\mathrm{L,2}$.
$\beta_\mathrm{BHL}$ is the mass-accretion efficiency predicted by BHL. It is determined from Equation \ref{p2:eq:BHL} with $\alpha_\mathrm{BHL} = 0.75$.
$(\dot{a}/a)_\mathrm{bin}$ is the rate of change in the orbital separation of the binary system. It is derived from Equation \ref{p2:eq:adot}.	
$(\dot{a}/a)_\mathrm{dyn}$ is the rate of change in the orbital separation of the binary system. It is derived dynamically from the simulation. 
Note that for the corotating systems we cannot make a direct comparison between $(\dot{a}/a)_\mathrm{bin}$ and $(\dot{a}/a)_\mathrm{dyn}$ because in the first case the change in the orbital separation is computed from the change in the orbital angular momentum, whereas when measured dynamically the rate of change in $a$ includes the change of angular momentum due to the spin of the donor star. 
}
\end{table*}

\subsection{Terminal velocity of the wind}
\begin{figure}
\centering
\includegraphics[width=\hsize]{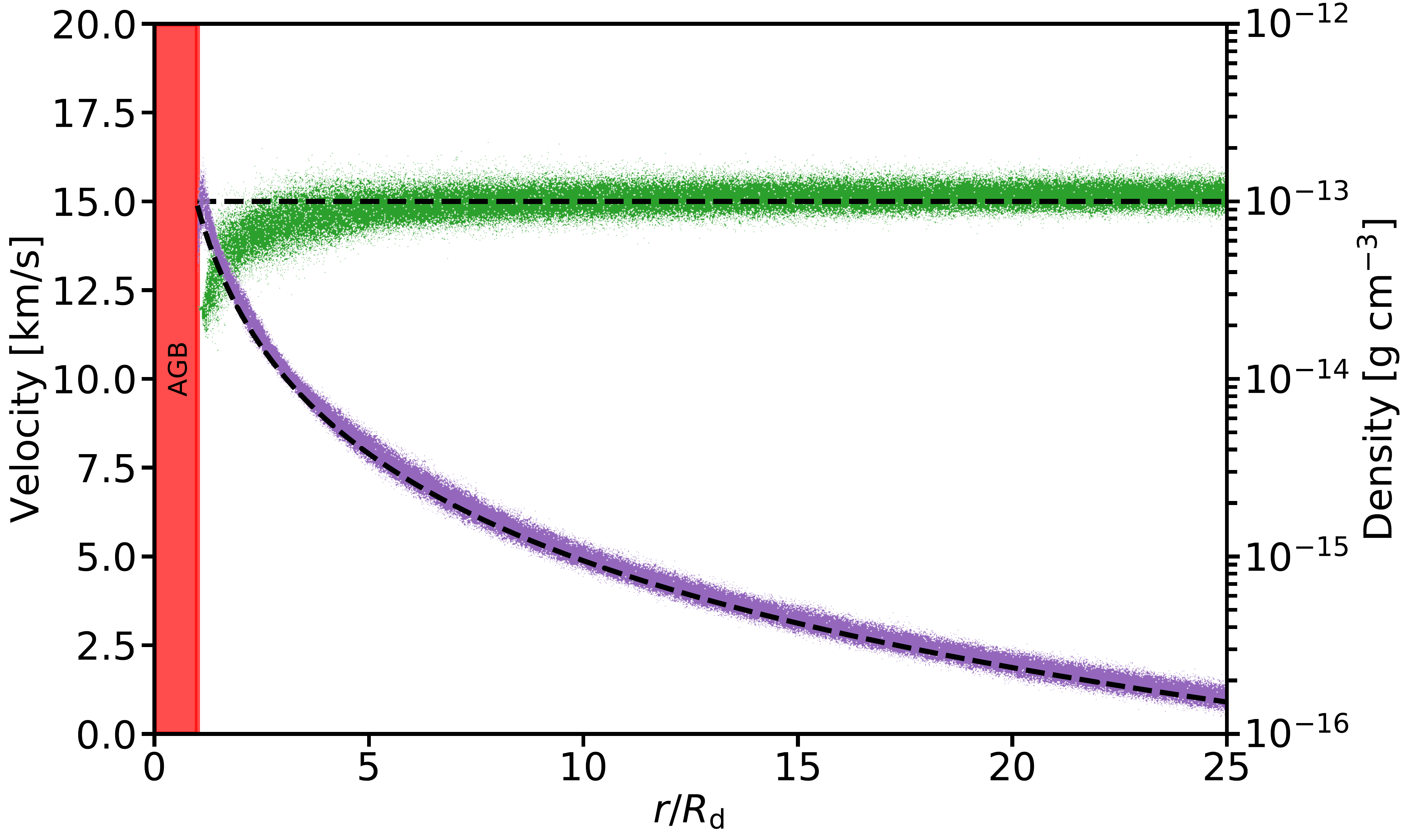}
\caption{Velocity and density profiles of the SPH gas particles leaving a single star with $v_\mathrm{init} = 12$ km s$^{-1}$. The top dashed line corresponds to 
a constant velocity of 15.1 km s$^{-1}$. The lower dashed line is the corresponding analytical density in a steady state. The red rectangle delimits the inner boundary of the SPH particles around the donor star.}
\label{p2:fig:single_profile}
\end{figure}

As mentioned in section \ref{p2:sec:numerical_method}, in this work the acceleration of the wind only balances the gravitational deceleration by the AGB star. 
For this reason, the terminal velocity of the wind will be different from that with which the wind is injected.
In Paper I, where we imposed a constant velocity on the wind particles with a predefined terminal velocity, we only explored velocities as low as 10 km s$^{-1}$ for numerical reasons\footnote{
Modelling systems with lower wind velocity is computationally more expensive since it takes longer time to reach the steady state.}. 
Here we explore lower values of the initial wind velocities. 
We determine the terminal velocities that can be achieved by performing simulations of single AGB stars. 
In these models the star is allowed to eject matter for 40 yr in order to reach stable conditions.
The average terminal velocities are measured at a distance of 25$R_\mathrm{d}$, and averaged over the last 1.5 yr. 
Figure \ref{p2:fig:single_profile} shows the velocity and density profiles of the wind as a function of distance for a single star with stellar parameters as shown in Table \ref{p2:table:donor_star} and an initial velocity of the wind of $v_\mathrm{init} = 12$ km s$^{-1}$. Because of gas pressure, the wind accelerates to an average terminal velocity $v_{\infty} = 15.1$ km s$^{-1}$. 
For $v_\mathrm{init} = 5$ km s$^{-1}$, we find that the average $v_{\infty} = 10.8$ km s$^{-1}$ and for $v_\mathrm{init} = 1$ km s$^{-1}$, $v_{\infty} = 6.0$ km s$^{-1}$. 
The injection of particles with such low initial wind velocities allows the wind to achieve terminal velocities which lie within the observed range as given by \cite{VassiliadisWoods1993} for the winds of AGB stars.
We describe the results of our binary simulations in terms of the terminal velocity achieved by the wind of a single star with the same initial velocity as assumed for the donor star. 

\subsection{Morphology of the outflow}

In the first part of this section we present the results on the outflow morphologies for the non-rotating models.
In the second part, we discuss the differences between the models in which corotation of the AGB donor is assumed and the corresponding models in which rotation is neglected. 

\subsubsection{Non-rotating models}\label{p2:sec:morpho_nr}

\begin{figure*}
\centering
\includegraphics[width=0.99\hsize]{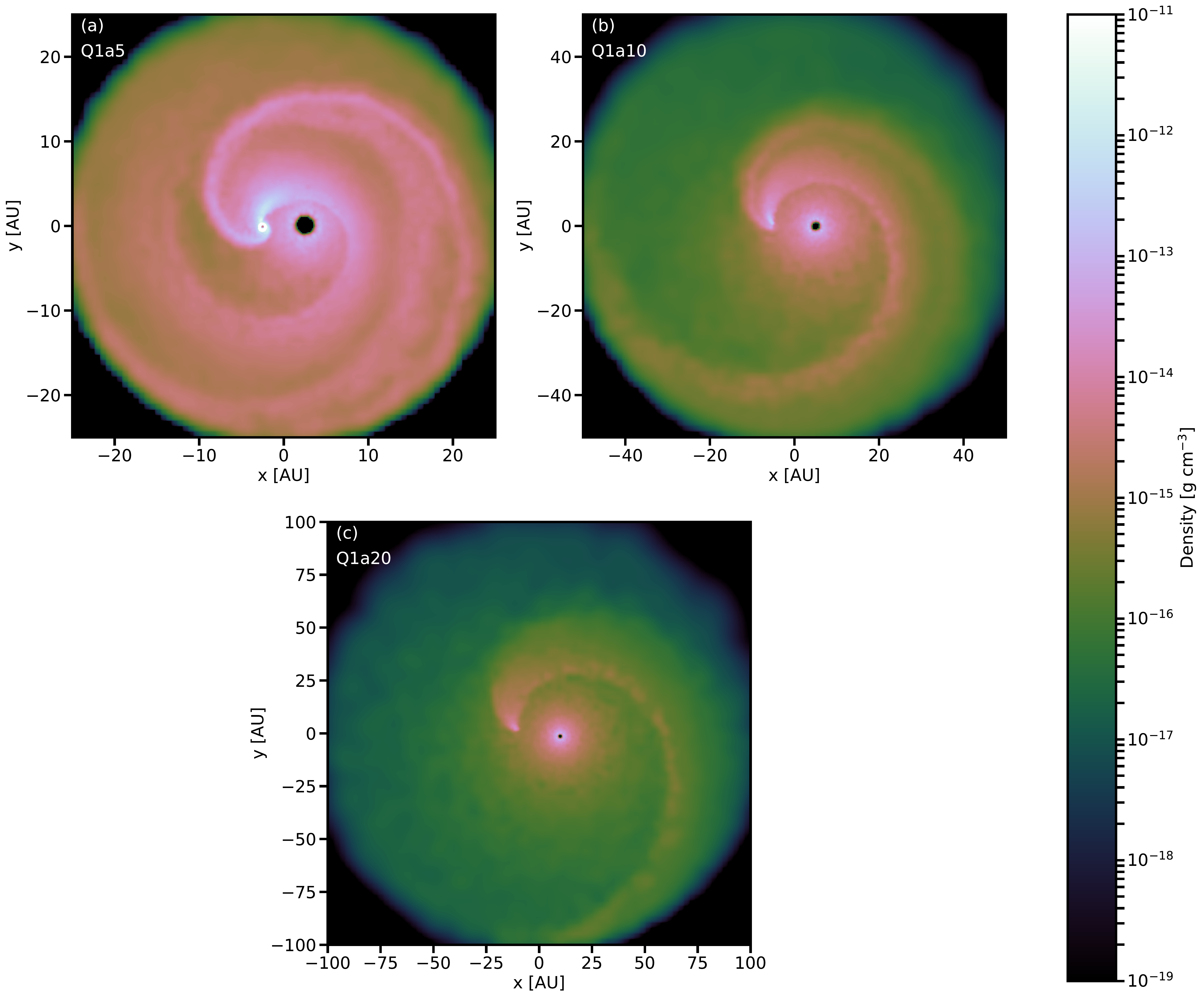}
\caption{Gas density in the orbital plane for models Q1a5, Q1a10 and Q1a15 (the same mass ratio, but different orbital separations). The companion star is located on the left and the AGB donor on the right. Each figure is scaled to display the full simulation, i. e. a sphere with radius equal to 5$a$, and shows the situation after 9.5 orbital periods.}
\label{p2:fig:density_a}
\end{figure*}

\begin{figure*}
\centering
\includegraphics[width=0.99\hsize]{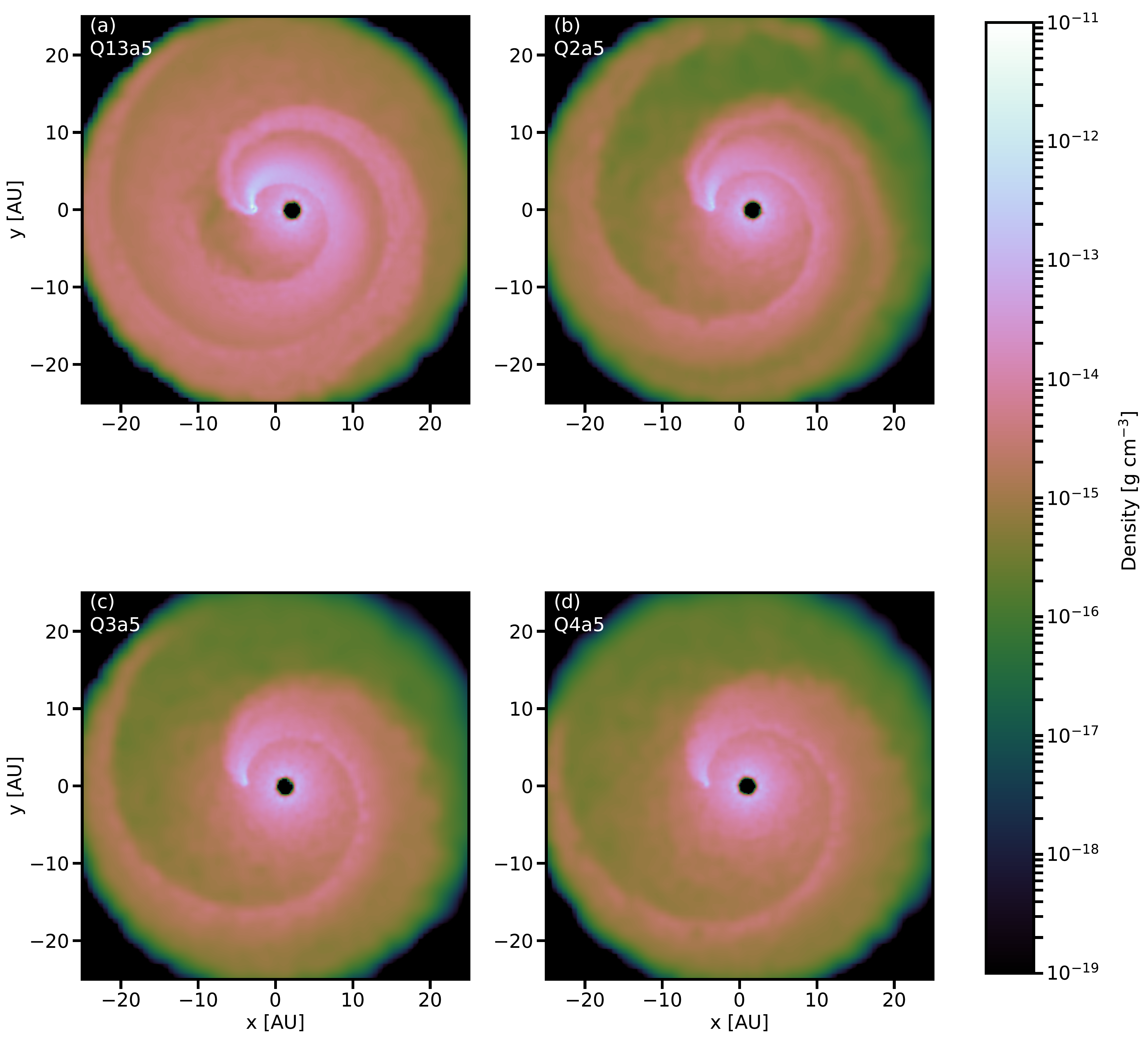}
\caption{Same as Figure \ref{p2:fig:density_a}, but for models with the same orbital separation ($a=5$ AU) and different mass ratios.}
\label{p2:fig:density_q}
\end{figure*}

\begin{figure*}
\centering
\includegraphics[width=0.99\hsize]{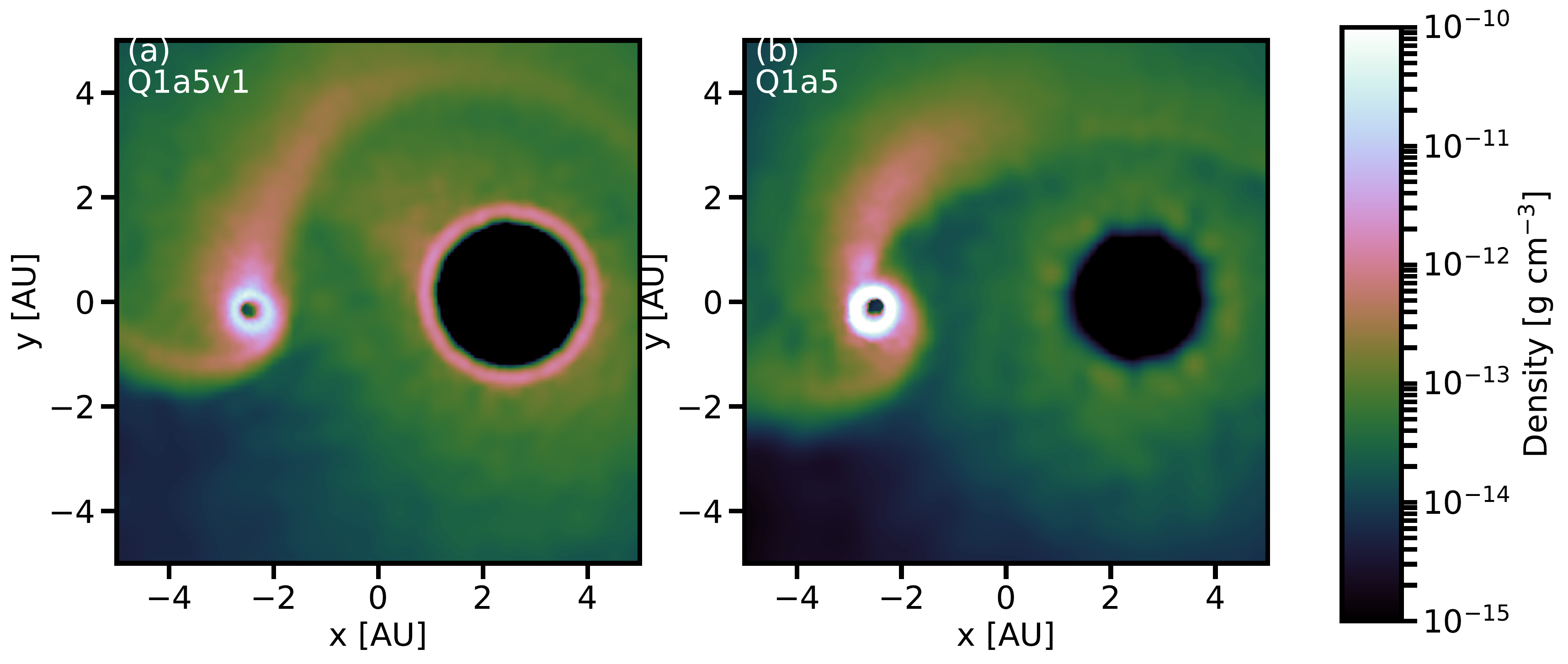}
\caption{Gas density on the orbital plane for for models with the same orbital separation and mass ratio, ($a=5$ AU, $q=1$), but different initial wind velocities (1 km s$^{-1}$ and 12 km s$^{-1}$).
Notice that for better appreciation in this figure the density scale is different from the previous figures.
Also notice that for model Q1a5 (right) this figure is a close-up version of Fig. \ref{p2:fig:density_q}a.}
\label{p2:fig:density_v}
\end{figure*}

The outflow morphologies in the simulations with a non-rotating donor star are very similar to those found in Paper I and they are in agreement with the results of similar works \citep[e.g.][]{mastrodemos, Liu+2017}. 
In general, for all the models we observe two spiral arms wrapped around the binary system. 
These spiral arms delimit the accretion wake behind the companion star. 
In the following we will describe the observed morphologies as a function of the orbital separation (Figure \ref{p2:fig:density_a}), the mass ratio (Figure \ref{p2:fig:density_q}) and the wind velocity (Figure \ref{p2:fig:density_v}).  

Figure \ref{p2:fig:density_a} shows the gas density profile in the orbital plane for models Q1a5, Q1a10 and Q1a20, which have the same mass ratio ($q=1$) but different orbital separations ($a=5, 10$, and 20 AU), at the same orbital phase. 
We observe that the complexity of the morphology decreases with increasing orbital separations, which suggests that as the orbital separation increases, the interaction between the companion star and the wind becomes less strong. 
In relatively close systems, the gas density in the spiral arms is higher than in wider systems. 
Also, the spiral arms which delimit the accretion wake behind the companion star are less tightly wound around the binary for wider systems. 
One more feature that varies in these models is the opening angle, $\theta$, of the accretion wake. 
We find that for close orbital separations, $\theta$ is large and it decreases when the orbital separation increases. 
This occurs because $\theta$ is a function of the Mach number, $\mathcal{M}$, i.e. $\theta = \sin^{-1} \mathcal{M}^{-1}$ (see Appendix \ref{p2:ap:mach}). 
We find that for the model with $a = 5$ AU, $\mathcal{M} \approx 2 $ in the vicinity of the accretor while the Mach number increases to $\mathcal{M} \approx 4.5 $ and $\mathcal{M} \approx 5$ for the systems with $a = 10$ AU and $a = 20$ respectively. 
In addition, we find that for model Q1a5 an accretion disk is formed, which is not observed in the wider models. 
However, as is extensively discussed in Paper I, since the radius of the sink particle is large, the disk may not be able to form in these cases. 

The same trend in the morphology of the outflow with orbital separation is found for systems with larger mass ratios. 
For close orbital separations a stronger interaction between the gas and the stars occurs, as shown by the presence of an accretion disk around the companion star and dense tightly wound spiral patterns around the binary. 

In order to illustrate the dependence of the morphology on the mass ratio, Figure \ref{p2:fig:density_q} shows the density profile in the orbital plane for models Q13a5, Q2a5, Q3a5 and Q4a5 (the same orbital separation, but different mass ratio). 
This can be compared with model Q1a5 (Figure \ref{p2:fig:density_a}a), which has the same orbital separation. However, these models do not show an accretion disk around the companion star, for the reasons explained above. 
Overall the gas densities are similar and all models have two arms tightly wound around the binary. 
The main differences occur in the accretion wake behind the companion star, which has lower density and a smaller opening angle as the mass ratio increases.
Similar to \cite{Liu+2017}\footnote{Note that their definition of mass ratio is $Q=1/q$.}, we find that for higher mass ratios the spiral arm is less tightly wound. 
These differences in the geometry of the outflow can be explained by the lower gravity of the companion star. 
When the companion star is more massive, its gravitational attraction is stronger and so is the spiral shock formed in the vicinity of the accretor. 
At the same time, the opening angle of the spiral shock becomes wider. 
This is explained by the dependence on the Mach number as discussed above.
As the mass ratio increases from $q=1$ to $q=4$, the Mach number near the companion increases from $\mathcal{M} \approx 2.5$ to $\mathcal{M} \approx 5$. 
On the other hand, as the mass ratio increases, the density along the inner spiral arm decreases. 
As will be shown in Section \ref{p2:sec:eta}, a lower density in the accretion wake implies a weaker torque on the binary system which affects the amount of angular momentum exchanged between the orbit and the gas. 

In general we observe that as the orbital separation increases, the structures in the morphology of the outflow become less prominent and less complex. 
This suggests that the interaction between the gas and the stars decreases as a function of distance. 
A similar trend was found in Paper I, where the free parameter in the simulations was the velocity of the wind. 
In the models discussed above the terminal velocity of the wind is fixed and we vary the orbital separation, which is equivalent to changing the orbital velocity of the system. 
As a result, the ratio $v_{\infty}/v_\mathrm{orb}$ of the models changes. 
We find very similar flow morphologies as in Paper I when comparing models with the same $v_{\infty}/v_\mathrm{orb}$.
Furthermore, we confirm that the relationship found in Paper I on the morphology as a function of $v_{\infty}/v_\mathrm{orb}$ also holds for different mass ratios. 

Finally, in order to study how the wind velocity influences the morphology of the outflow, in Figure \ref{p2:fig:density_v} we compare two models with a mass ratio $q=1$, a small orbital separation ($a = 5$ AU), and initial wind velocities of 1 km s$^{-1}$ and 12 km s$^{-1}$. 
Since the sound speed close to the donor star is about 6 km/s, in the first model (Q1a5v1) the wind leaves the donor star in the subsonic regime, whereas for higher-velocity model (Q1a5) the wind is  
supersonic when it leaves the donor star. 
In contrast to Figures \ref{p2:fig:density_a} and \ref{p2:fig:density_q}, only the inner region of the simulation is shown in Figure \ref{p2:fig:density_v}.
In model Q1a5 (Fig. \ref{p2:fig:density_v}b) the wind leaving the donor remains spherically symmetric until it reaches the accretion shock and the inner spiral arm. 
However, when the initial wind velocity is subsonic (model Q1a5v1, Fig. \ref{p2:fig:density_v}a), the outflow in the vicinity of the donor star's surface is clearly asymmetric and partly focussed into a broad stream that is deflected behind the companion's accretion wake. 
The deflection is related to the non-rotation of the donor star, as will be discussed in Section \ref{p2:sec:morpho_r}.
We note that due to the small orbital separation in these models, the surface of the donor star and the wind launch region will be somewhat deformed by tides from the companion. 
This is not taken into account in our simulations, but is likely to enhance the asymmetry in the inner part of the outflow seen in Fig. \ref{p2:fig:density_v}b.
On larger scales than shown in Fig. \ref{p2:fig:density_v}, both models show a spiral pattern similar to that observed in Fig. \ref{p2:fig:density_a}a. 
However, as expected because of the low wind velocity, the density in the spiral arms of model Q1a5v1 is higher than in model Q1a5.

\subsubsection{Rotating models}\label{p2:sec:morpho_r}

\begin{figure*}
\centering
\includegraphics[width=0.9\hsize]{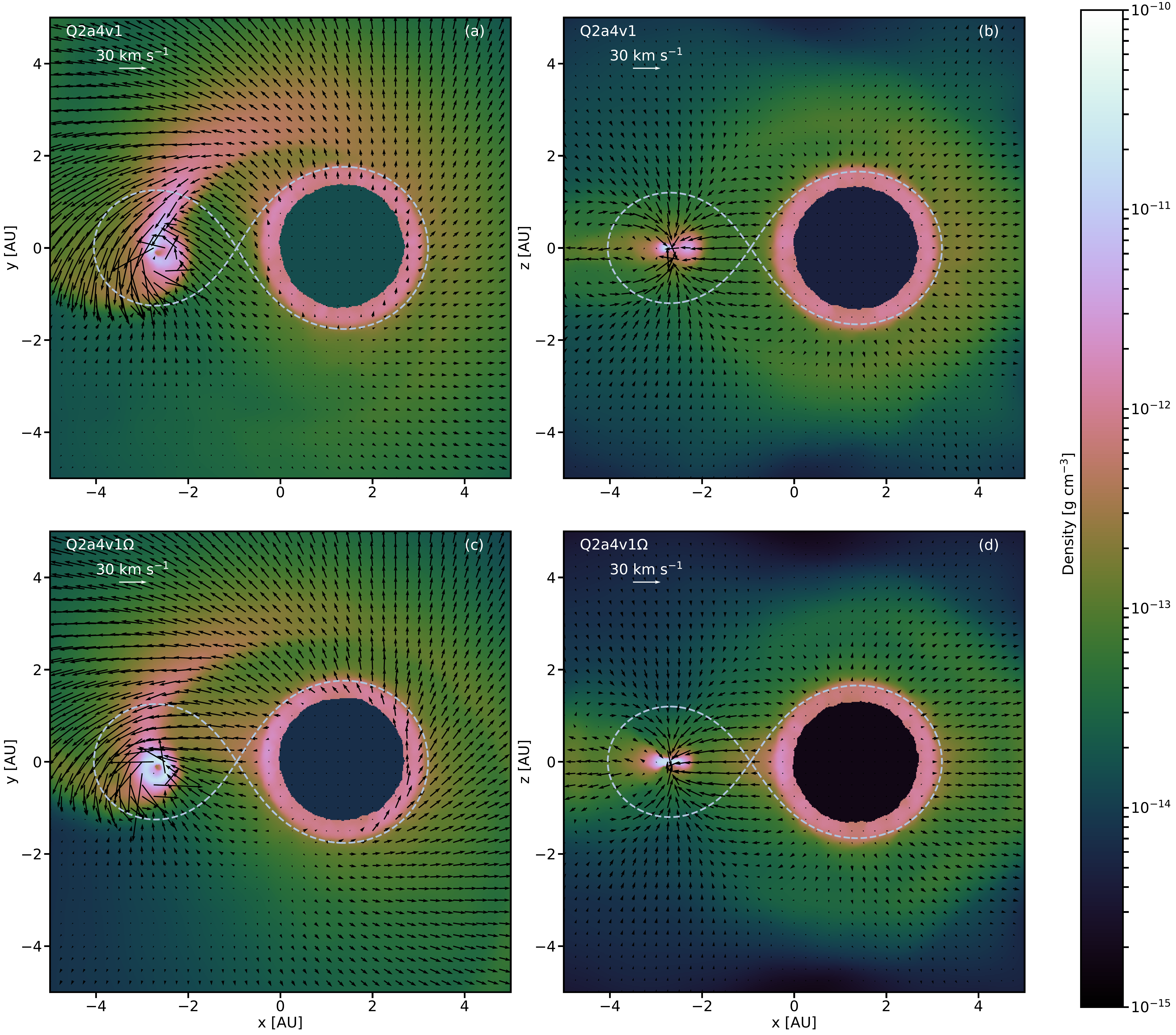}
\caption{Density of the gas for models Q2a4v1 and Q2a4v1$\Omega$ after 9.5 orbital periods. The images on the left show the density and velocity field in the inertial frame in the orbital plane ($z=0$). The images on the right show the same quantities but on the $y=0$ plane. The dashed lines correspond to the Roche-lobes of the stars.}
\label{p2:fig:velocity_field_spin}
\end{figure*}

In the models where the AGB donor is in corotation with the binary, we find two different regimes in the geometry of the outflow. 
This difference appears to be related to the initial velocity of the wind. 
For models Q2a5$\Omega$ and Q4a5$\Omega$, where $v_\mathrm{init} = 12$ km s$^{-1}$, the outflows are very similar to their non-rotating counterparts (Q2a5 and Q4a5; see Figure \ref{p2:fig:density_q}). 
For model Q1a5$\Omega$ (not shown), we observe a similar morphology to that of Q1a5 (Figure \ref{p2:fig:density_a}a). Both systems have an accretion disk and two spiral arms tightly wound around the binary. However, the inner spiral arm for model Q1a5 is slightly denser than the one for model Q1a5$\Omega$.  

On the other hand, when we compare the corotating models with small orbital separations (4 AU and 5 AU) and low initial velocities of the wind ($v_\mathrm{init} = 1$ km s$^{-1}$ and $v_\mathrm{init} = 5$ km s$^{-1}$) with their non-rotating counterparts, we note clear differences in the morphologies of the outflow. 
We note that at these orbital separations, the companion star is located at a position where the gas has not yet reached its terminal velocity and that the donor star is very close to filling its Roche lobe ($R_\mathrm{d}/R_\mathrm{L,1} = 0.86$ for Q2a4v1$\Omega$ and $R_\mathrm{d}/R_\mathrm{L,1} = 0.69$ for Q2a5v5$\Omega$). 
Interestingly, the geometries in these corotating models resemble the WRLOF morphology found in simulations performed in the corotating frame \citep[e. g.][]{val-borro, shazrene, shazrene1, Chen+2017}.
In Section \ref{p2:disc:morphology} we compare the results of our corotating models Q2a4v1$\Omega$ and Q2a5v5$\Omega$ to the results of simulations performed in the corotating frame. 

Figure \ref{p2:fig:velocity_field_spin} shows the gas density on the planes $z=0$ (left) and  $y=0$ (right) for models Q2a4v1 (top) and Q2a4v1$\Omega$ (bottom). 
The velocity field in the inertial frame is plotted in the same figure. 
In spite of the accretion disk around the companion star which is a common feature in both models, there are many dissimilarities between the models. 
A clear difference that can be observed in the orbital plane, $z=0$ (Figures \ref{p2:fig:velocity_field_spin}a and \ref{p2:fig:velocity_field_spin}c), is the high gas density along the inner spiral arm for model Q2a4v1 (Figure \ref{p2:fig:velocity_field_spin}a; $x\approx -2.5$ AU, $0$ AU $\lesssim y \lesssim 0.1$ AU), which is substantially lower in the corotating model Q2a4v1$\Omega$ (Fig. \ref{p2:fig:velocity_field_spin}c). 
Also, the opening angle of the inner arm is larger in the non-rotating model and wraps around the donor star at closer distances than in the corotating model.   
Another clear difference is the stream of gas flowing from the donor star to the companion star in model Q2a4v1$\Omega$, which can be observed in both planes ($z=0$ and $y=0$). 
In the orbital plane, this stream is seen to flow from the inner Lagrangian point, $L_1$, at about $x \approx -0.49$ AU to $x \approx -2.5$ AU, where it encounters the accretion wake of the companion star. 
The presence of the stream between the two stars in Fig. \ref{p2:fig:velocity_field_spin}c, which resembles RLOF, creates the impression that the effect of the rotating donor is to focus the wind in the direction of the accretor.
Conversely, model Q2a4v1 does not show this stream of gas moving from $L_1$ towards the companion star, but instead the gas moving from the AGB star towards the companion is deflected behind this star, where it is captured by the accretion wake and then falls onto the accretion disk. 
However, unlike model Q2a4v1, in model Q2a4v1$\Omega$ the gas appears to pass through the accretion wake once it reaches it. 
This might be due to the fact that for Q2a4v1 the density in the accretion wake is higher than for the rotating model, and is able to trap the gas more easily when it reaches it.
In addition, from the flow in the $y=0$ plane (Figs. \ref{p2:fig:velocity_field_spin}b and \ref{p2:fig:velocity_field_spin}d), we observe a region in model Q2a4v1$\Omega$ where material is moving away from the binary ($ -3$ AU  $ \lesssim x \lesssim -5$ AU and $-1$ AU $\lesssim$ $z \lesssim 1$ AU). 
Finally, in model Q1a4v1, the material moving away is confined to a narrower band around the $z=0$ plane, whereas the gas at $z \gtrsim 0.5$ AU and $z \lesssim  -0.5$ AU is observed to be moving towards the companion star.  

The morphology of model Q2a5v5$\Omega$ (not shown) is somewhat similar to that of model Q2a4v1$\Omega$ with gas being focused in the direction of the companion star. 
We find a stream of gas flowing through the inner Lagrangian point and moving towards the inner spiral arm. Similar to Q1a4v1$\Omega$, the inner spiral arm is less dense than its non-rotating analogue. 
The main difference we find with model Q1a4v1$\Omega$ is that the region in the $y=0$ plane where gas moves away from the binary is narrower in the $z=0$ plane giving the impression that less material escapes the system. 

\subsection{Angular-momentum loss} \label{p2:sec:eta}

The loss of angular momentum for non-conservative mass transfer in binaries can be parameterised as\footnote{
Note that Eq. \ref{p2:eq:eta} ignores the rotational velocity of the  secondary star, which could accrete angular momentum from the wind and be spun up. However, since we model the accretor star as a sink we cannot account for the real angular momentum that is transferred (see Paper I for a discussion).
}:
\begin{equation}
\label{p2:eq:eta}
\dot{J} = \eta a^{2} \Omega_\mathrm{bin}(1-\beta)\dot{M}_\mathrm{d} + \dot{J}_\mathrm{spin}.
\end{equation}
The first term corresponds to the change in the orbital angular momentum, $\dot{J}_\mathrm{orb}$, where $\beta = -\dot{M}_\mathrm{a}/\dot{M}_\mathrm{d}$ is the fraction of ejected mass which is accreted by the companion star, $\dot{M}_\mathrm{d} <0$ is the mass-loss rate, and $\eta$ is the specific angular momentum of the material lost in units of the orbital angular momentum of the system per reduced mass, $J/\mu=a^2\Omega_\mathrm{bin}$. 
The second term in Equation \ref{p2:eq:eta} corresponds to the angular momentum lost due to rotation of the donor star. If the donor star is assumed to be non-rotating, this term is zero. 

In a similar fashion to Paper I, we take the values of the mass and angular momentum lost from the system as those taken away by gas particles which cross a boundary of $3a$ from the centre of mass of the system. In order to reduce statistical fluctuations, we average the angular-momentum loss over intervals of one orbital period from the time the simulation reaches the steady state, which occurs after about four orbits \citep[cf.][Section 3.5]{Saladino+2018a}. 

In the simulations in which rotation is included, the numerical value measured for $\dot{J}$ is the combination of the angular momentum removed from the orbit and the angular momentum taken from the spin of the donor star, $\dot{J}_\mathrm{spin}$, which we cannot disentangle. 
In order to apply Equation \ref{p2:eq:eta} and derive the value of $\eta$, $\dot{J}_\mathrm{spin}$ is assumed to be given by the analytical equation for isotropic mass loss from a spherical surface:
\begin{equation}
\label{p2:eq:J_spin}
\dot{J}_\mathrm{spin} = \frac{2}{3} R_\mathrm{d}^2\dot{M}_\mathrm{d}\Omega_\mathrm{spin},
\end{equation}
where $\Omega_\mathrm{spin} = \Omega_\mathrm{bin}$ in the case of corotation. 
In order to verify that the analytical value for $\dot{J}_\mathrm{spin}$ given by Equation \ref{p2:eq:J_spin} accurately represents the spin angular-momentum loss, we perform a numerical simulation of a single star with the properties of Table \ref{p2:table:donor_star}, rotating at $\Omega_\mathrm{spin} = 2.76 \times 10^{-8}$ s$^{-1}$. 
We find that the numerical value obtained for $\dot{J}_\mathrm{spin}$ agrees with Equation \ref{p2:eq:J_spin} within 2\%. 
For this reason, for corotating systems in which the flow morphology is not strongly modified by rotation (see Section \ref{p2:sec:am_r}), we expect Equation \ref{p2:eq:eta} to yield similar $\eta$
values as for their non-rotating analogues. 

In the following we present the results for the angular-momentum loss of our models in terms of the parameter $\eta$. 
First we discuss the models in which the donor star is assumed to be non-rotating ($\dot{J}_\mathrm{spin} = 0$) and in the second part we present the results for the angular-momentum loss of the corotating models. 

\subsubsection{Non-rotating models}\label{p2:sec:am_nr}

\begin{figure}
\centering
\includegraphics[width=\hsize]{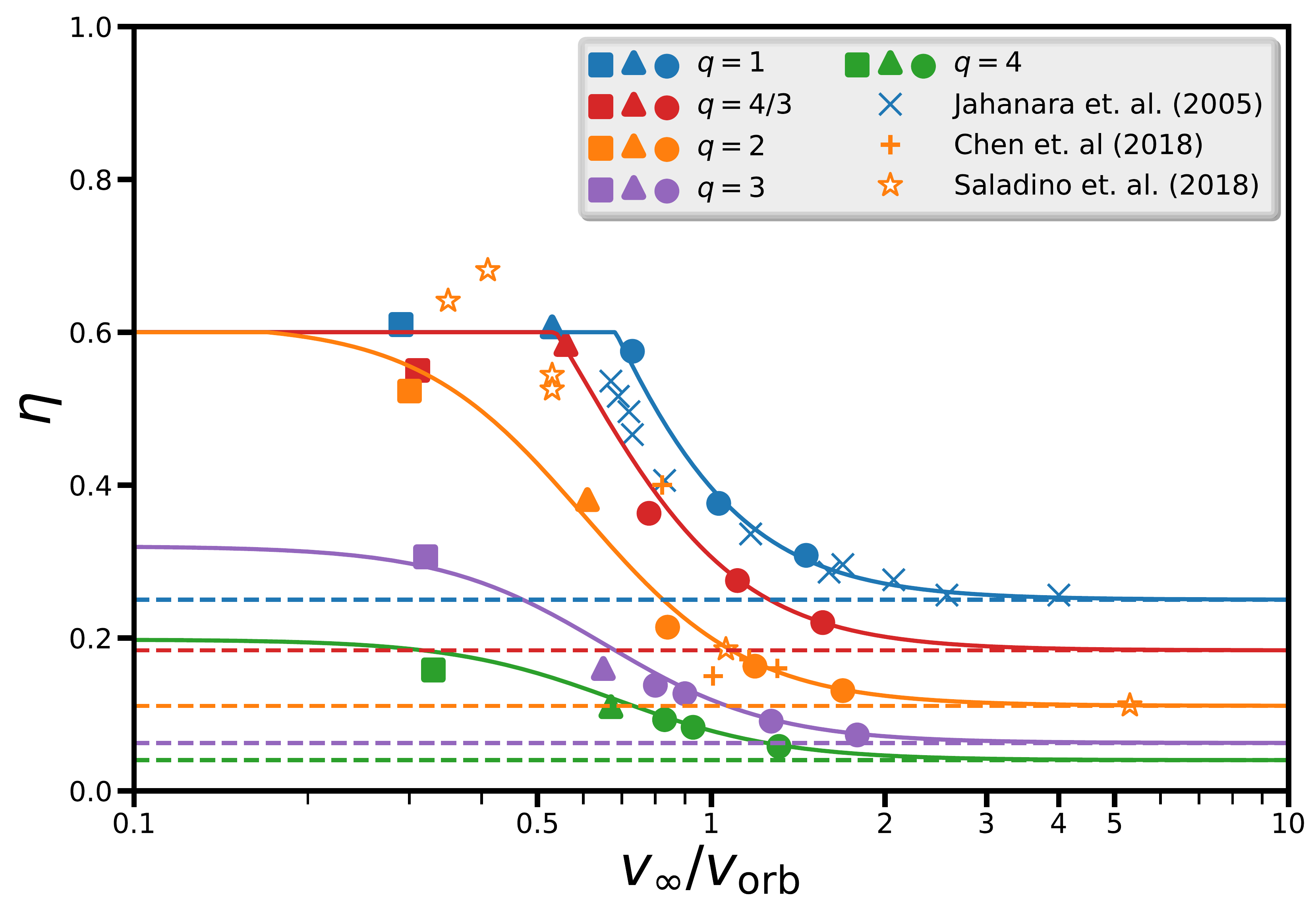}
\caption{Orbital angular-momentum loss in units of $J/\mu$ as a function of $v_{\infty}/v_\mathrm{orb}$ for models assumed to be non-rotating. Colours represent equivalent mass ratios. Dashed lines show the value of $\eta_\mathrm{iso}$ for different mass ratios. Filled circles correspond to models in which $v_{\infty} = 15.1$ km s$^{-1}$, triangles for models where $v_{\infty} = 10.9$ km s$^{-1}$ and squares for models where $v_{\infty} = 6$ km s$^{-1}$. The solid lines correspond to Equation \ref{p2:eq:eta_fit} for different mass ratios.}
\label{p2:fig:eta}
\end{figure}


Figure \ref{p2:fig:eta} shows the specific angular momentum lost, $\eta$, in units of $J/\mu$ as a function of the velocity ratio $v_\infty/v_\mathrm{orb}$ for the different non-rotating models studied in this paper. 
Colours represent equivalent mass ratios.  
This figure only shows those systems with non-rotating donor stars. 
When compared to the results of Paper I (open stars in Figure \ref{p2:fig:eta}), where angular-momentum loss was studied only for $q=2$, we observe a similar trend of the $\eta$ values as a function of $v_\infty/v_\mathrm{orb}$. 
We emphasize that in Paper I the parameter we varied was $v_\infty$, whereas here $a$ is varied.  
For high velocity ratios, the values of $\eta$ approach the isotropic wind\footnote{This corresponds to the fast or Jeans mode where mass lost from the star occurs in the form of spherical symmetric isotropic wind, and the angular-momentum loss is the specific angular-momentum loss of the donor star in its relative orbit.} values given by:
\begin{equation}
\label{p2:eq:eta_iso}
\eta_\mathrm{iso} = \frac{1}{(1+q)^2},
\end{equation}
which are shown as dashed lines in Figure \ref{p2:fig:eta}. 
On the other hand, when $v_\infty/v_\mathrm{orb}$ decreases, the angular-momentum loss of the models in enhanced. 
A notable difference with Paper I is that the $\eta$ values found in this work for low $v_\mathrm{init}$ are smaller than those found in Paper I at the same $v_\infty/v_\mathrm{orb}$. 
However, we should keep in mind that in this work there are a number of differences in the numerical method compared to Paper I. 
The stellar radius and effective temperature of the donor star are larger in this work. 
This leads to a different temperature profile in the outflow, which may influence the interaction between the gas and the stars. 
In addition, the mechanism of injection of the wind particles is different from the one used in Paper I. 
This may affect the angular momentum transfer, in particular when the companion star is located at a position where the wind is still being accelerated. 

From Figure \ref{p2:fig:eta}, we also observe that as the mass ratio increases, the angular-momentum loss decreases, for similar values of $v_\infty/v_\mathrm{orb}$.
This behaviour is expected according to the results discussed in Section \ref{p2:sec:morpho_nr}. 
Given the lower densities behind the accretion wake for larger mass ratios, the torque exerted by the wake on the binary is weaker, resulting in a smaller exchange of angular momentum. 
An interesting feature we find is that for low $v_\infty/v_\mathrm{orb}$ and $q\lesssim 2$, $\eta$ appears to level off at a maximum value of about 0.6. 
However, when $q\gtrsim 3$, $\eta$ never reaches such high values.  
Finally, although for most of our simulations $\eta$ seems to grow monotonically with decreasing $v_\infty/v_\mathrm{orb}$ until reaching its maximum for the lowest $v_\infty/v_\mathrm{orb}$, this does not seem to be the case for the $q=4/3$ models. For these models the maximum of $\eta$ occurs for $v_\infty/v_\mathrm{orb} \approx 0.6$ and it decreases for the lowest $v_\infty/v_\mathrm{orb}$. 

In order to apply our results in binary population synthesis simulations, the angular-momentum loss needs to be expressed as a function of the physical parameters of the systems. 
In Paper I, we provide an analytic relation for the angular-momentum loss as a function of $v_\infty/v_\mathrm{orb}$ but independent of $q$. 
However, we find that this relation only matches our current results for high $v_\infty/v_\mathrm{orb}$ and mass ratios between 1 and 2. 
For this reason, we provide a new function which describes our results as a function of the mass ratio and the ratio between the terminal velocity of the wind and the system's orbital velocity:

\begin{equation}
\eta(q, v_{\infty}/v_\mathrm{orb}) = \min \left(\frac{1}{c_1 + (c_2 v_\infty/v_\mathrm{orb})^3} + \eta_\mathrm{iso}, \  0.6\right), 
\label{p2:eq:eta_fit}
\end{equation}
where
\begin{equation}
c_1 = \max(q, \ 0.6 \ q^{1.7}),
\end{equation}
\begin{equation}
c_2 = 1.5 + 0.3 q,
\end{equation}
and
$\eta_\mathrm{iso}$ is given by Equation \ref{p2:eq:eta_iso}.

This fit is overplotted in Figure \ref{p2:fig:eta} for several values of the mass ratio, corresponding to our simulations. 
For comparison, we also plot the results for the angular-momentum loss from \citet[blue crosses]{jahanara2}, where the wind is accelerated in a similar fashion as in this work, and the results of \citet[orange plus signs]{rochester}, who used a different method to model the stellar wind. For both works, we apply Equation \ref{p2:eq:eta} to the published results to correct for the spin angular momentum due to corotation of the system. 

\subsubsection{Rotating models}\label{p2:sec:am_r}

\begin{figure}
\centering
\includegraphics[width=\hsize]{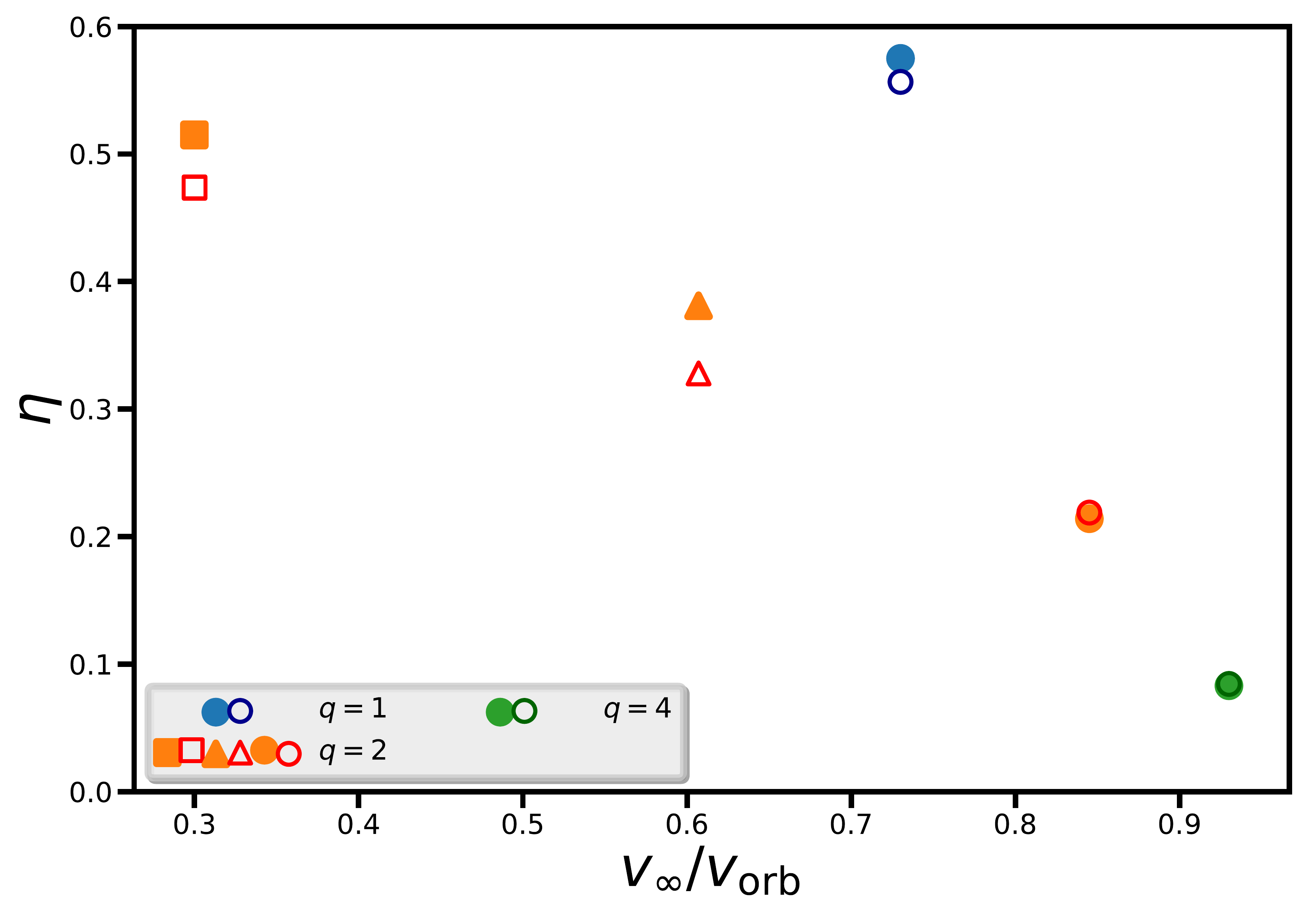}
\caption{Orbital angular-momentum loss in terms of $\eta$ for models in corotation with the binary and their non-rotating analogues. Colours represent equivalent mass ratios. Circles
correspond to models where the initial velocity was set as $v_\mathrm{init} = 12$ km s$^{-1}$ ($v_\infty = 15.1$ km s$^{-1}$) and squares to models where $v_\mathrm{init} = 1$ km s$^{-1}$ ($v_\infty = 6$ km s$^{-1}$). Open shapes are the corresponding corotating models.}
\label{p2:fig:eta_spin}
\end{figure}

Figure \ref{p2:fig:eta_spin} shows the orbital angular-momentum loss, as measured by the parameter $\eta$, as a function of the ratio $v_\infty/v_\mathrm{orb}$ for the corotating models (open shapes) with their non-rotating counterparts (filled shapes). 
We observe that for $v_{\infty}/v_\mathrm{orb} \gtrsim 0.8$ the orbital-angular-momentum loss is equal for the rotating and non-rotating models. 
This means that the extra angular-momentum loss measured in the corotating models indeed comes from the spin of the donor star 
and is correctly described by Equation \ref{p2:eq:J_spin}. 
However for low $v_{\infty}/v_\mathrm{orb}$, the orbital angular-momentum loss in the corotating models is lower than that measured for the non-rotating models.  
The difference is small for model Q1a5$\Omega$, and somewhat larger for Q2a4v1$\Omega$ ($\approx$ 8\%) and Q2a5v5$\Omega$ ($\approx$ 13\%). 
We note that $\eta$ measures the exchange of angular momentum between the escaping gas and the binary orbit. 
The fact that $\eta$ is independent of rotation for larger $v_{\infty}/v_\mathrm{orb}$ is consistent with the fact that we find no differences in the morphology between rotating and non-rotating models at these velocities (see Section \ref{p2:sec:morpho_r}). 
The lower angular momentum exchange in models Q2a4v1$\Omega$ and Q2a5v5$\Omega$ compared to Q2a4v1 and Q2a5v5 respectively,  could be due to a smaller torque exerted by the accretion wake on the binary in the rotating models, given the lower density in the inner spiral arms for these models compared to their non-rotating analogues. 

\subsection{Mass-accretion efficiency}\label{p2:sec:beta}

\begin{figure}
\centering
\includegraphics[width=\hsize]{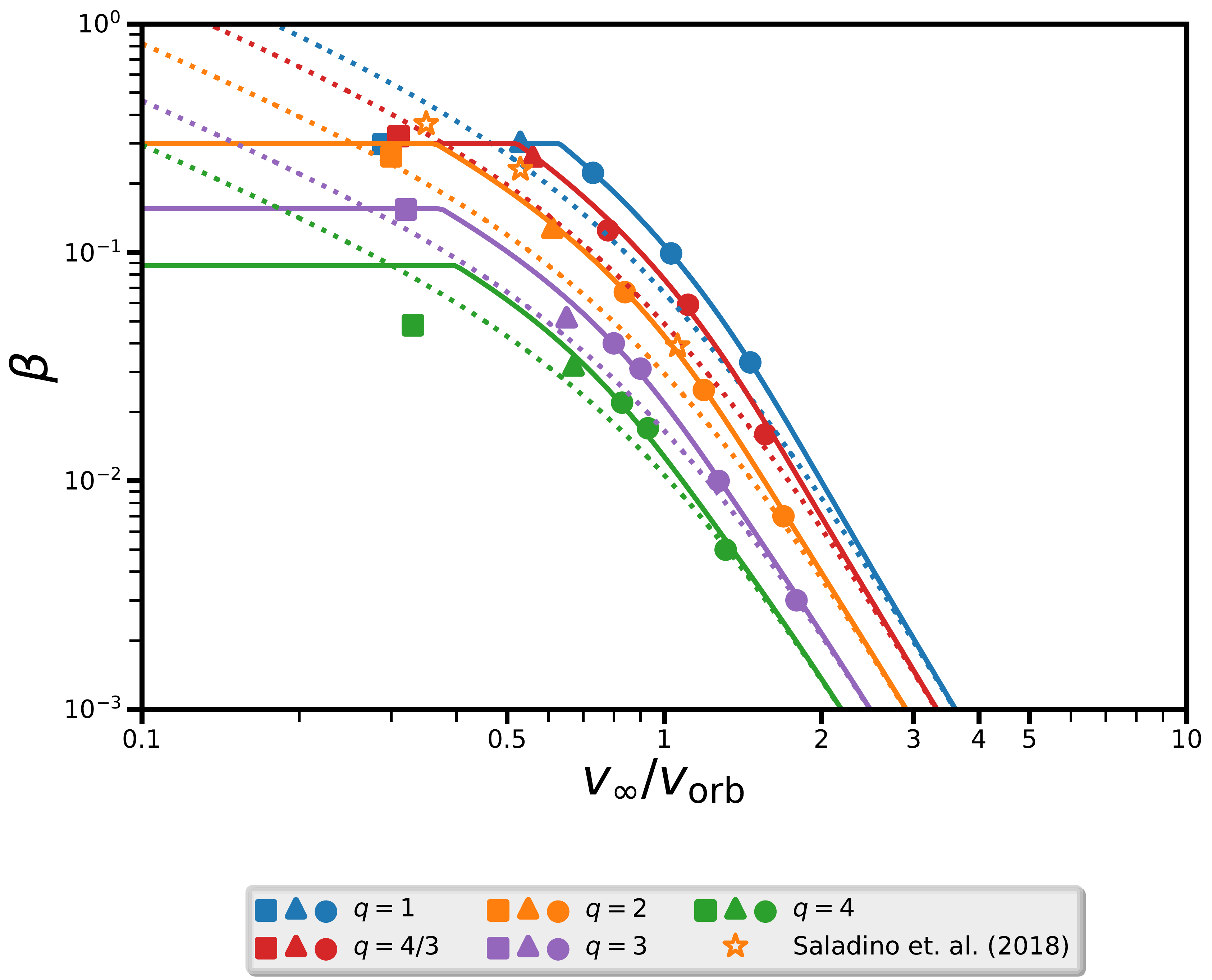}
\caption{Mass-accretion efficiency for the non-rotating models. Colours represent equivalent mass ratios. Circles
correspond to models where $v_\infty = 15.1$ km s$^{-1}$, triangles to models where $v_\infty = 10.9$ km s$^{-1}$ and squares to models where $v_\infty = 6$ km s$^{-1}$. The dotted lines correspond to the values predicted by BHL as given by Equation \ref{p2:eq:BHL} assuming $\alpha_{BHL} = 0.75$. Continuous lines correspond to the fit given by Equation \ref{p2:eq:beta_fit}.}.
\label{p2:fig:beta}
\end{figure}

\begin{figure}
\centering
\includegraphics[width=\hsize]{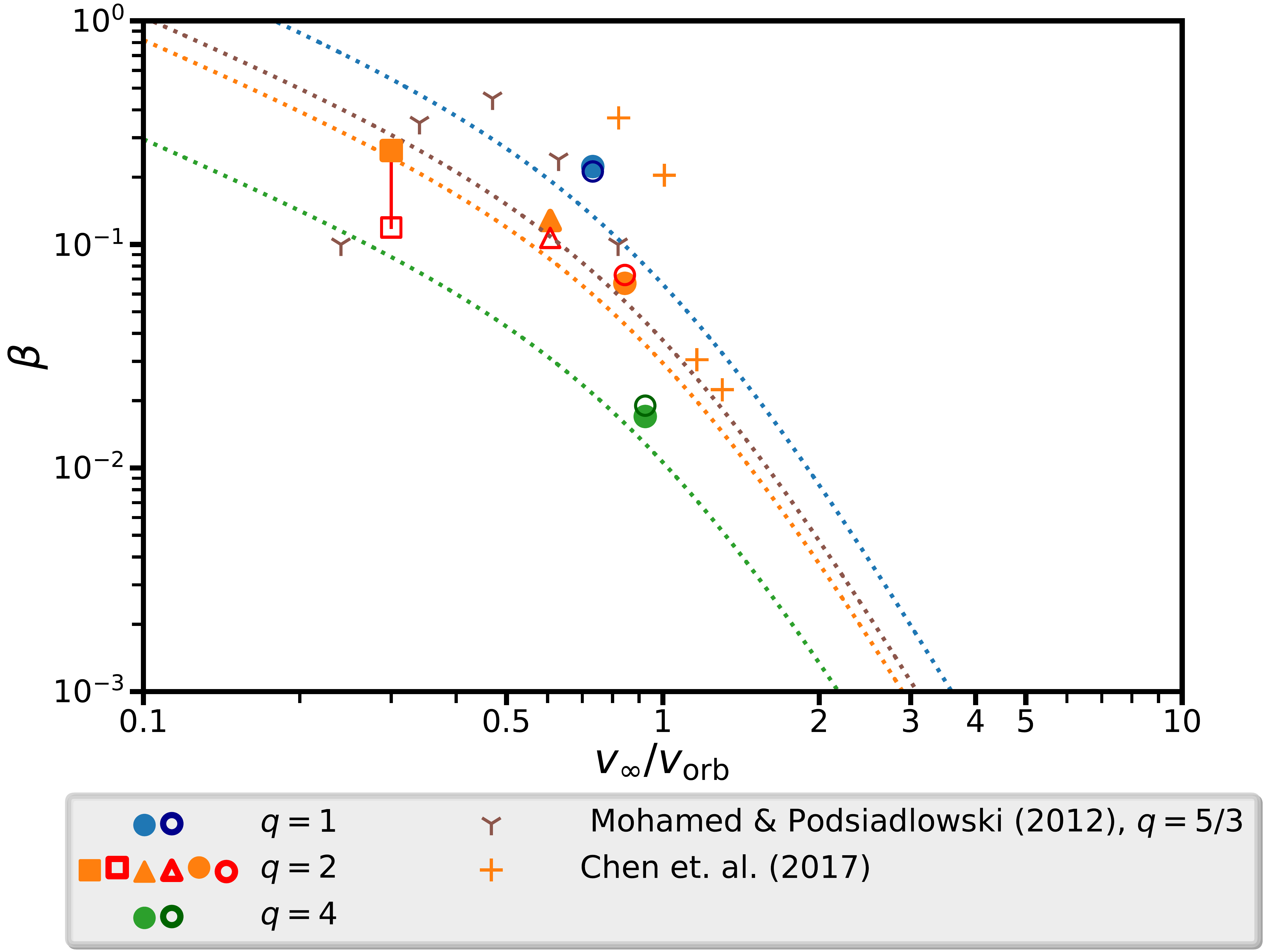}
\caption{Similar to Figure \ref{p2:fig:beta} but for the rotating models.}
\label{p2:fig:beta_rot}
\end{figure}

The BHL approximation provides an estimate of the mass-accretion efficiency onto a body moving at given speed in a high-velocity wind assuming supersonic flow. 
Under this approximation, in the case of a binary system in which one of the stars is losing mass via winds, the companion star will accrete a fraction of the mass lost by the donor given by:
\begin{equation}
\label{p2:eq:BHL}
\beta_\mathrm{BHL} = \frac{\alpha_\mathrm{BHL}}{(1+q)^2} \frac{v_\mathrm{orb}^4}{v_\mathrm{w}(v_\mathrm{w}^2 + v_\mathrm{orb}^2)^{3/2}},
\end{equation}
where $\alpha_\mathrm{BHL}$ is the efficiency parameter, which has a value between 0.5 and 1, and $(v_\mathrm{w}^2 + v_\mathrm{orb}^2)^{1/2}$ is the relative wind velocity seen by the accretor \citep{theuns2}. 

Considering the assumptions made in the canonical BHL approximation and the fact that AGB winds are very slow, 
a more suitable way to estimate the mass-accretion efficiencies is to measure them directly from detailed hydrodynamical simulations. 
However, we emphasize that since we cannot resolve the surface of the companion star computationally, our numerical results for the mass-accretion efficiency should be taken as upper limits. 
As discussed in Paper I, rather than measuring the mass flux that crosses the boundary of the sink particle, a more reliable measurement for the accretion rate onto the companion star is derived from the sum of the mass accretion rate on the sink particle and the rate at which mass is added to the accretion disk. 
For this reason, and following the method used in Paper I, we adopt as the mass-accretion efficiency the net flux crossing a shell centred on the companion star with radius equal to 0.4 $R_\mathrm{L, 2}$. 
Similar to the angular-momentum loss, to avoid statistical fluctuations, we measure the average mass-accretion efficiency over several orbital periods once the system has reached the steady state. 

\subsubsection{Non-rotating models}\label{p2:sec:beta_nr}

Figure \ref{p2:fig:beta} shows the fraction of mass accreted $\beta = \dot{M}_{0.4R_\mathrm{L,2}}/|\dot{M}_\mathrm{d}|$ as a function of $v_{\infty}/v_\mathrm{orb}$. 
The dotted lines correspond to the predicted BHL value (Equation \ref{p2:fig:beta}) with $\alpha_\mathrm{BHL} = 0.75$ \citep[as][]{carlo1}. 
Similar to Paper I, for individual mass ratios, we find that the smaller the $v_{\infty}/v_\mathrm{orb}$ ratio, the higher the mass-accretion efficiency we measure.
In general we notice that for different mass ratios and $v_{\infty}/v_\mathrm{orb} > 0.5$, the accretion efficiency measured in our models is always larger than the BHL prediction (see Table \ref{p2:table:results}). 
The maximum difference between the BHL approximation and our results occurs for models Q1a5 and Q13a5v5, where we find that the value of $\beta$ from the hydrodynamical models is about 1.6 times larger than the BHL prediction. 
Only when $v_{\infty}/v_\mathrm{orb}$ is very high the accretion efficiency approaches the BHL approximation. 
In the cases where this occurs (see Section \ref{p2:sec:morpho_nr}) the morphologies of the outflow are less complex suggesting less interaction between the wind and the companion star. 
On the other hand, for models with $v_\mathrm{init} = 1$ km s$^{-1}$ there is no clear trend in the accretion efficiencies we measure, which can be either higher or lower than the $\beta$ values predicted by the BHL approximation. 
The maximum accretion efficiencies we obtain never exceed about 30\%, in contrast with the accretion efficiencies of Paper I, where for low velocity ratios, we found values up to about 40\%. 
We should note that the stellar parameters used in this study differ from Paper I, and that the current numerical resolution  is much lower than the resolution used in Paper I, which could be influencing our results (see Sect. \ref{p2:sec:dis_beta}).

In order to implement our results in a binary population synthesis code, we construct a function that describes the behaviour found from our hydrodynamical simulations for the mass-accretion efficiency. Similar to the angular-momentum loss, we fit our results as a function of mass ratio and the ratio $v_{\infty}/v_\mathrm{orb}$. A relation that describes our numerical results is given by:

\begin{equation}
\label{p2:eq:beta_fit}
\beta(q, v_\infty/v_\mathrm{orb}) = \min(\alpha\beta_\mathrm{BHL}, \ \beta_\mathrm{max}),
\end{equation}
where:
\begin{equation}
\alpha = 0.75 + \frac{1}{k_1 + (k_2 v_\infty/v_\mathrm{orb})^{5}}
\end{equation}
and
\begin{itemize}
\item $k_1 = 1.7 + 0.3 q$,
\item $k_2 = 0.5 + 0.2 q$,
\item $\beta_\mathrm{max} = \min(0.3, 1.4 \ q^{-2})$,
\end{itemize}
and $\beta_\mathrm{BHL}$ is given by Equation \ref{p2:eq:BHL} with $\alpha_\mathrm{BHL} = 1$. This relation is shown in Figure \ref{p2:fig:beta} for different mass ratios.

\subsubsection{Rotating models}\label{p2:sec:beta_r}

Figure \ref{p2:fig:beta_rot} shows the mass accretion efficiencies for the models assumed to be in corotation. 
We find that the mass-accretion efficiency for models Q2a5$\Omega$ and Q4a5$\Omega$ is slightly larger (by a factor of about 1.1) than for their non-rotating counterparts, while for Q1a5$\Omega$ and Q2a5v5$\Omega$ it is slightly smaller. 
However, for model Q2a4v1$\Omega$, where the injection velocity of the wind is $v_\mathrm{init} = 1$ km s$^{-1}$ and the geometry of the outflow resembles WRLOF, we find that the accretion efficiency is less than half the value of its non-rotating analogue (see Table \ref{p2:table:results}). 
Although we may attribute this low mass-accretion efficiency to material crossing the low density accretion wake of the companion star (see Section \ref{p2:sec:morpho_r}), this appears to be in contrast to the SPH simulations performed by \cite{shazrene1} in which they find high accretion efficiencies when mass transfer occurs via WRLOF. 
For comparison, in Figure \ref{p2:fig:beta_rot}, we plot the mass accretion efficiencies from \cite{carlo1} based on the hydrodynamical simulations of \cite{shazrene_thesis} as a function of $v_{\infty}/v_\mathrm{orb}$ (brown Y shapes). 
We compute $v_\mathrm{orb}$ from the orbital periods provided in \cite{carlo1} and we take $v_{\infty} = 4$ km s$^{-1}$ as in \cite{shazrene}. 
In the same figure, we also show the mass-accretion efficiencies from the hydrodynamical models of \cite{Chen+2017} (orange plus signs). 
We observe that in both works the trend of the mass-accretion efficiency as a function of $v_{\infty}/v_\mathrm{orb}$ is similar to the one we find, i.e. $\beta$ decreases with increasing $v_{\infty}/v_\mathrm{orb}$. 
However, for $v_{\infty}/v_\mathrm{orb}$ in the range between 0.5 and 1, both \cite{shazrene} and \cite{Chen+2017} find much larger values than predicted by BHL. 
This differs from the values we find in the same velocity range, where our $\beta$ values are only slightly larger than BHL (see Section \ref{p2:sec:discussion} for a discussion). 
On the other hand, for $v_{\infty}/v_\mathrm{orb} < 0.5$, the $\beta$ values found by \cite{shazrene} decrease again and, similar to our results, for the lowest velocity ratio $\beta$ is much lower than the BHL expected value, even though WRLOF occurs.  

\subsection{Change in orbital separation}\label{p2:sec:adot}
\begin{figure}
\centering
\includegraphics[width=\hsize]{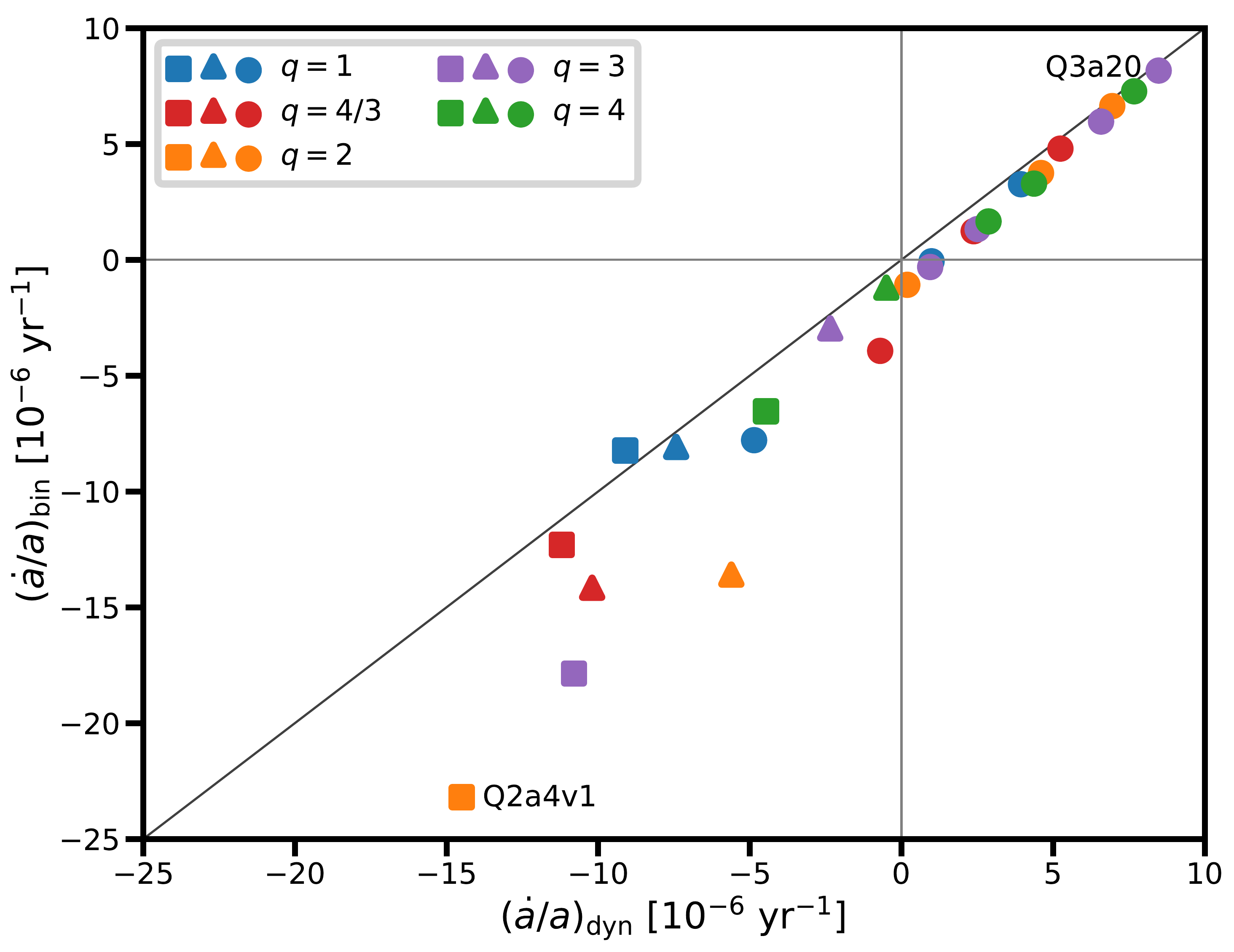}
\caption{Evolution of the orbital separation, $\dot{a}/a$, measured with two different methods. 
The values on the $x$-axis are measured directly from the evolution of the orbits of the stars. 
The values on the $y$-axis are obtained by substituting the parameters $\eta$ and $\beta$ measured from the gas dynamics in equation \ref{p2:eq:adot}.}
\label{p2:fig:adot}
\end{figure}

Due to the loss of angular momentum via stellar winds, the orbital separation of the binary will be affected. 
The rate of change of $a$ can be derived from angular momentum conservation (see Paper I):
\begin{equation}
\label{p2:eq:adot}
\frac{\dot{a}}{a} = -2 \frac{\dot{M_\mathrm{d}}}{M_\mathrm{d}} \left[1 - \beta q - \eta(1-\beta)(1+q) - (1-\beta) \frac{q}{2(1+q)}\right],
\end{equation}
where the parameters $\eta$ and $\beta$ are computed from detailed hydrodynamical simulations. We should note that when there is efficient spin-orbit coupling by tidal interactions, another term should be added to Equation \ref{p2:eq:adot} to take this into account. 

Another method to compute the change in $a$ is to measure it dynamically from our numerical simulations. 
To compute it in this way, we use the Newtonian description of the motion of two masses under their mutual gravitational field (see Appendix \ref{p2:ap:3}). From our numerical simulations we know the position $\mathbf{r}$ and the relative velocity of the stars $\mathbf{v}$ at any time. 
From these we compute the orbital energy and angular momentum per reduced mass, and by approximating the orbits as Keplerian, we compute instantaneous values of $a$ and $e$. We then calculate $\dot{a}$ as the average rate of change of $a$ over the last orbits. 
We find that $e$ is not exactly zero, but its value shows a sinusoidal variation for all models. 
However this variation is so small ($\sim 10^{-5}$) that it can be neglected. 

Figure \ref{p2:fig:adot} shows the change in orbital separation calculated in the two different ways described above for the non-rotating models. 
On the $x$-axis we plot the values of $\dot{a}/a$ measured directly from the simulations (Eqs. \ref{p2:eq:energy1} and \ref{p2:eq:energy2})
and on the $y$-axis we show the values computed by substituting the measured values of $\eta$ and $\beta$ (Table \ref{p2:table:results}) in Equation \ref{p2:eq:adot}. 
For most systems in which the orbit expands both values agree quite well. 
In these binaries, the wind leaves the system in an almost spherically symmetric way. 
However, in systems with shrinking orbits we find a substantial difference between the two $\dot{a}$ values. 
The discrepancy appears to be correlated with the speed at which the orbit shrinks, and in most cases the systems shrink slower if measured directly from the simulations. 
We note that the cases where the discrepancy is large correspond to models where the interaction between the gas and the stars is strong (squares and triangles in Figure \ref{p2:fig:adot}), i.e. where more angular momentum is transferred from the binary orbit to the gas. 
This means that in these cases, if angular momentum is not conserved precisely, an error in this quantity may build up. 

For a TreeSPH code like \textsc{Fi} angular momentum is not conserved exactly \citep{wijnen1}. 
For this reason, we compute how well angular momentum is conserved for our models. 
In order to do so, we compute the total initial angular momentum, 
$J_\mathrm{tot, init} = J_\mathrm{stars, init} + J_\mathrm{gas, init}$, where $ J_\mathrm{stars, init}$ is the initial orbital angular momentum of the binary and $J_\mathrm{gas, init}$ is the angular momentum of the gas which is initially in the simulation. 
We also compute the total angular momentum at the moment we stop the simulations, $J_\mathrm{tot, end} = J_\mathrm{stars, end} + J_\mathrm{gas, end} + J_\mathrm{acc} + J_\mathrm{esc}$, where $J_\mathrm{stars, end}$ is the orbital angular momentum at the last timestep of our simulation,  $J_\mathrm{gas, end}$ is the angular momentum of the gas at the end of the simulation, $J_\mathrm{acc}$ is the cumulative angular momentum of the gas accreted by the sink particle and $J_\mathrm{esc}$ the cumulative angular momentum of the gas we removed during the simulation. 
We express the error, $\delta J$, as the difference $J_\mathrm{tot, end} - J_\mathrm{tot, init}$ in units of the total initial angular momentum per unit of time. We find that in most of our models angular momentum is conserved fairly well.  
For model Q3a20, where the difference between the two methods to compute $\dot{a}/a$ is only 4\%, we find that $\delta J = 4.46 \times 10^{-7} J_\mathrm{tot, init}$ yr$^{-1}$.
 However, for model Q2a4v1 where the discrepancy in $\dot{a}/a$ is $\approx 37\%$, angular momentum is conserved up to $2.47 \times 10^{-5} J_\mathrm{tot, init}$ per year. 
Since the error in the angular momentum of the stars and the gas is a fraction of the total error computed and we cannot separate the errors, it is not possible to precisely correct the $\dot{a}/a$ values computed either dynamically or from the parameters $\beta$ and $\eta$.
However, we have verified by error propagation that both $(\dot{a}/a)_\mathrm{bin}$ and $(\dot{a}/a)_\mathrm{dyn}$ are consistent within the total error in the angular momentum budget for each system.

\section{Discussion}\label{p2:sec:discussion}

\subsection{On the morphology of the outflow: rotating vs non-rotating systems}\label{p2:disc:morphology}

Observations of low-mass binary systems interacting via winds show spiral patterns wrapped around the stars \citep[e.g.][]{Kim+2017, Ramstedt+2017}. 
Numerical simulations also produce these features in the outflow \cite[e.g.][]{mastrodemos, Liu+2017, Saladino+2018a}. 
In addition, some planetary nebulae, which are thought to be the final evolutionary stage of low- and intermediate-stars, show morphologies which cannot be explained by single-star evolution.
For these reasons, as is discussed in Paper I, understanding how the morphologies are influenced by the interacting binary can provide an insight into the mechanisms driving mass transfer between the stars.

In this work we find that the spin of the star plays a significant role in shaping the morphology of the outflow. 
For cases in which the angular velocity of the donor star is zero, we find similar morphologies to those in Paper I: two spiral arms surrounding the binary system and in some cases an accretion disk. 
Similar to Paper I, a strong interaction of the wind with the stars can be observed as an increasingly complex morphology in the outflow, that is, the morphology of the wind differs significantly from the spherical symmetric case. In addition, we find that the strength of the interaction depends on the orbital separation and mass ratio of the binary system. 
With increasing orbital separation the morphology of the outflow becomes less complex, and for large orbital separations the wind escapes the binary in a nearly isotropic way. 
Varying the orbital separation of the system results in similar geometries as a function of $v_{\infty}/v_\mathrm{orb}$ as when $v_{\infty}$ is varied for a fixed separation (as has been done in Paper I). 
In terms of the mass ratio, we find spiral patterns that become less prominent as the mass ratio is further from unity. Similar structures of which the complexity decreases as a function of the mass ratio are also found by \cite{Liu+2017}, who used a different EoS. 
These results on the morphology can help to constrain the orbital separations and mass ratios of binary observations.

Similar to the zero-spin models, in the rotating models we find that the geometry of the outflow becomes less complex with increasing $v_\mathrm{\infty}/v_{\mathrm{orb}}$, and more closely resembles the spherically symmetric wind case.
A similar result is observed in the hydrodynamical models in the corotating frame of \cite{Chen+2017}, where although a spiral outflow morphology is found for wide binaries, this becomes less prominent with increasing orbital separation (i.e. with increasing $v_{\infty}/v_\mathrm{orb}$). 
We find that the effect of rotation on the morphology of the wind is negligible when $v_{\infty}/v_\mathrm{orb} \gtrsim 0.7$ and that for these models the geometries of the outflow are similar to those observed when the star is not rotating. 
However, when the wind velocity is low ($v_\mathrm{init} = 1$ km s$^{-1}$ and $v_\mathrm{init} = 5$ km s$^{-1}$) and the system is relatively close, we find structures in the outflow which resemble WRLOF. 
The morphologies we find are similar to those observed by \cite{val-borro}, \cite{shazrene} and  \cite{Chen+2017} in the corotating frame. 
By comparing our results with these works, we notice that these geometries are generally found when the velocity of the wind is much lower than the orbital velocity of the binary. 
For instance, in their more recent hydrodynamical models of binary stars interacting via winds, \cite{Val-Borro+2017} do not find the prominent stream flowing between the stars discussed in \cite{val-borro}, where the adopted wind velocities were lower.

In the physical interpretation of \cite{shazrene1}, WRLOF is expected to occur when the dust formation region is larger than the Roche lobe of the star, confining material to the gravitational potential of the donor star and filling the Roche lobe. 
Material is then transferred to the companion star via the inner Lagrangian point. 
Our models do not include dust formation and the acceleration of the wind simply balances the gravity of the star. 
However, given that some acceleration via gas pressure still occurs, the velocity of the wind within the Roche lobe is very small (see Table \ref{p2:table:results}) allowing the wind material to fill the Roche lobe and part of the material to be transferred in a way that resembles RLOF. 
Similar to the models by \cite{val-borro} and \cite{shazrene}, we observe that the stream of gas is not exactly focused towards the companion star, but slightly deflected towards the accretion wake. 
Morphologies like those described here are also observed. 
For instance, recent observations by \citet{Bujarrabal+2018} of the symbiotic binary R Aqr show an arc of gas joining the two stars, which is attributed to an episode of mass transfer taking place in the system.

When we compare our models on spatial scales larger than about  5AU (as shown in Figure \ref{p2:fig:velocity_field_spin}) to figure 4 of \cite{Chen+2017}, we observe that generally the outflow geometries are very similar, with two differences.
In the inner region of the binary system our models show the flow to be very steady compared to theirs. 
A likely explanation for this difference is that in their models pulsations of the AGB star are included which make this region more dynamic.  
The second difference we note is that in some of their models, especially where the morphologies are similar to WRLOF, a circumbinary disk is found.
In our simulations we do not see such  a structure. 
However, we should keep in mind that in our models we remove the escaping material at a relatively short distance from the binary and that we do not include dust formation nor radiative transfer, which may be relevant for the formation of this disk. 
Longer numerical simulations which include this physics would be needed to check whether a circumbinary disk is formed. 

Finally, we note that the peculiarities in the geometries of the outflow discussed here are likely to have an impact on the evolution of the binary via differences in the mass-accretion efficiency and angular-momentum loss.
For instance, a denser accretion wake exerts a higher torque on the binary resulting in a larger exchange of angular momentum between the orbit of the stars and the gas.

\subsection{Angular-momentum loss and the impact on the orbit}

Recent hydrodynamical studies of angular-momentum loss during wind mass transfer have explored its dependence on only a few parameters. 
\cite{jahanara2} performed grid-based hydrodynamical models of binary systems undergoing mass loss via radiatively-driven winds, i.e., similar to this work the acceleration of the wind only balances the gravitational force of the donor star. 
In their models, the EoS used is adiabatic and the mass ratio is $q=1$, they varied the injection velocity of the wind. 
\cite{rochester} also performed grid-based hydrodynamical simulations which include pulsations of the AGB donor star, dust formation, cooling of the gas and radiative transfer. 
Except for one model, the mass ratio ($q=2$) was fixed and they studied the dependence of the angular-momentum loss as a function of the orbital separation of the binary. 
In Paper I we performed a few exploratory simulations which did not include acceleration of the wind, but included cooling of the gas. 
The works of \cite{jahanara2}, \cite{rochester} and Paper I show a very similar dependence of the angular-momentum loss on $v_{\infty}/v_\mathrm{orb}$. 
In this work, we have explored a larger grid of binary parameters and wind velocities and we include the possibility of corotation of the donor star with the binary. 
In the following we extend the comparison to our present work. 

\subsubsection{Non-rotating models}

Figure \ref{p2:fig:eta} shows the orbital angular-momentum loss, expressed in terms of $\eta$, as a function of $v_{\infty}/v_\mathrm{orb}$ for the non-rotating models. 
When we compare this to the results of \citet[their radiatively driven wind case]{jahanara2}, \cite{rochester} and Paper I, we see that all models show a good agreement for $v_{\infty}/v_\mathrm{orb}>1$, where $\eta$ tends towards the isotropic-wind value (see Equation \ref{p2:eq:eta_iso}).
For the cases where $v_{\infty}/v_\mathrm{orb}<1$, there is a clear trend of $\eta$ increasing with decreasing $v_{\infty}/v_\mathrm{orb}$. 
This behaviour is expected because the velocity of the wind is much lower than the relative orbital velocity of the stars, thus the wind has a stronger interaction with the companion star, which allows it to draw more angular momentum from the orbit. 
However, we notice that despite following the same trend, there is a slight discrepancy between Paper I and this work in the $\eta$ values found as a  function of $v_{\infty}/v_\mathrm{orb}$ .
We associate this difference to the different assumptions made for the acceleration of the wind, as well as the different stellar parameters of the donor star. 
The different acceleration mechanism can lead to close binaries interacting in a region where the wind has not reached the terminal velocity.
The difference in stellar masses and metallicity can influence the temperature profile of the gas, since in this work the effective temperature of the donor star is larger. 
These two effects influence the interaction between the wind and the binary. 
Furthermore, unlike what we anticipated in Paper I, our current models show that the values of $\eta$ converge to a maximum of $\eta \approx 0.6$ for low $v_{\infty}/v_\mathrm{orb}$ and $q \leq 2$.

\subsubsection{Rotating models}

For models in which rotation is included, we find two regimes for the measured $\eta$ values.
For models with $v_{\infty}/v_\mathrm{orb} \gtrsim 0.7$, where the spin of the star does not affect the geometry of the gas (see Section \ref{p2:sec:morpho_r}), we find that the orbital angular momentum carried away by the escaping gas is independent of the spin of the donor star, i.e. the values we obtain for $\eta$ as derived from Equation \ref{p2:eq:eta} are the same as in their non-rotating counterparts.
This implies that the terms $\dot{J}_\mathrm{orb}$ and $\dot{J}_\mathrm{spin}$ act independently and that the orbital evolution of the system can be predicted by considering separately the change in angular momentum due to mass lost from the system and the change in angular momentum due to tides, as usually assumed in binary population synthesis codes. 
However, for $v_{\infty}/v_\mathrm{orb} \lesssim 0.7$, the spin of the star modifies the outflow which also changes the interaction between the gas and the stars compared to the non-rotating models. 
Consequently, we find different $\eta$ values for the corotating and non-rotating models. 
This implies that for cases in which the outflow is strongly modified by the spin of the star, detailed simulations including rotation of the donor are needed to correctly predict the amount of orbital angular-momentum loss and therefore the evolution of the orbit. 

In the numerical models performed by \cite{rochester} in the corotating frame the parameter that is varied is the orbital separation of the system. 
They explain why synchronisation of the donor star with the binary is expected only for some of the orbital separations they explore. 
In order to investigate the cases in which the donor star is not rotating, they provide a prescription in which they simply subtract the contribution of the spin of the star from the total angular-momentum loss they measure (similar to what we do in this work).
However, because of the results discussed above, the prescription they use to study the zero-spin cases for the AGB star can only be applied when the rotation of the donor does not modify the outflow of the gas. 
In addition, due to the nature of their numerical models, they find that gas gains angular momentum as it moves radially away from the mass-losing star, which they attribute to numerical errors due to the intrinsic viscosity of Eulerian codes. 
This is not observed in our numerical models since SPH codes are better at conserving angular momentum than grid-based codes.

As mentioned in Section \ref{p2:sec:adot}, from Equation \ref{p2:eq:eta} we can predict the rate of change of the orbital separation of the binary system due to orbital angular-momentum loss. 
However, if tidal interactions are effective in the binary, additional terms need to be taken into account in that equation. 
As shown in Appendix \ref{p2:appendix:rot}, for orbital separations up to $\approx 10$ AU, tidal friction is so strong that spin-orbit coupling is expected at some point during the AGB phase of the donor star. 
Furthermore, even when tidal interactions are not strong enough for synchronisation they can effectively spin up the donor star (see Figure \ref{p2:fig:omegas}b) for orbital separations between 10-20 AU.
In order to keep the donor star in corotation with the binary via tidal effects, angular momentum needs to be taken from the orbit of the binary.
This means that for some of the orbital separations considered in this work tidal evolution plays an important role, implying that additional shrinking of the orbit will take place. 
 
\subsection{Mass-accretion efficiency}\label{p2:sec:dis_beta}

Observations of the progeny of AGB binary systems show enhanced chemical abundances in AGB nucleosynthesis elements, which sometimes cannot be explained by the standard BHL formalism \citep[e.g. the chemical abundances of observed CEMP-$s$ stars][]{carlo2, carlo3}.
Since the assumptions made for the flow in the BHL formalism usually do not hold for the winds of AGB stars, hydrodynamical models are needed to obtain better estimates of the mass-accretion efficiency onto the companion star.
Several previous hydrodynamical studies have focused on computing this quantity (e.g. \cite{theuns2, val-borro, shazrene1, Liu+2017, Val-Borro+2017}). 
However, from the numerical point of view this is not an easy task.  
Since the radius of the companion star cannot be resolved numerically, these works have used a variety of methods to model the accretion process onto the surface of the star. 
In addition, some works have shown  that the mass-accretion efficiency depends on the EoS used and on the numerical resolution of the simulation \citep{theuns2, Saladino+2018a}. 
For these reasons comparing results from different studies is not straightforward. 
However, given some resemblances in the trends of mass-accretion efficiency as a function of mass ratio and $v_{\infty}/v_\mathrm{orb}$, we can attempt to do so. 

Figure \ref{p2:fig:beta} shows that in agreement with Paper I for the non-rotating models, the mass accretion efficiency can be described as a function of $v_{\infty}/v_\mathrm{orb}$, where for large $v_{\infty}/v_\mathrm{orb}$ values $\beta$ approaches the BHL approximation. 
Similar to \cite{Liu+2017}, we find that $\beta$ depends on the mass ratio, with higher accretion efficiencies for binary stars with more comparable masses. 
\cite{Liu+2017} provide an analytical expression that describes their numerical results for the mass-accretion efficiency as a function of the mass ratio.  
Since in their models a constant orbital separation of 3 AU is adopted and the stellar parameters used were different than in our models, we cannot compare our $\beta$ results directly with their analytical expression. 
However, even for our models with the closest separation ($a =5$ AU), we find $\beta$ values that are much larger than predicted by their analytical expression. 
In their models they use an adiabatic EoS to model the thermodynamics of the gas. 
As shown by \cite{theuns2} and \cite{Saladino+2018a} for their models with $q =2$ and $a = 3$ AU, an adiabatic EoS usually leads to lower accretion efficiencies than predicted by BHL.
For these reasons, the fit we provide in Equation \ref{p2:eq:beta_fit} is more appropriate for binaries interacting via winds.  

We can compare the $\beta$ values of our corotating models with \cite{shazrene} and \cite{Chen+2017}, who made similar assumptions for the spin of the donor star. 
Similar to the non-rotating models, these models show a dependence on the ratio $v_{\infty}/v_\mathrm{orb}$. 
However, in our models $\beta$ is only a factor of $\sim 1.6$ larger than in the standard BHL model, whereas in both \cite{shazrene} and \cite{Chen+2017} the difference is much larger (a factor of 2 or more). 
Although each of these works uses a different numerical method to measure the accretion rate onto the companion star, this is unlikely to explain the differences between the results. 
For instance, the accretion region in \cite{Chen+2017} is smaller than the sink particle assumed in this work, which should result in a lower accretion rate. 
\cite{shazrene_thesis} finds that the smooth method used in \cite{shazrene} to model the accretion process produces lower accretion efficiencies than immediately removing particles from the simulation after they enter the sink region, as we do. 
On the other hand, we note that the SPH resolution chosen in our models is quite low in order to make the simulations computationally efficient. 
As shown in Paper I, a resolution like the one adopted in our models can underestimate the accretion efficiency by a small factor compared to a higher resolution.
It seems more likely that the different accretion efficiency found in our work compared to \cite{shazrene} and \cite{Chen+2017} originates from the differences in the assumed input physics. 
Both \cite{shazrene} and \cite{Chen+2017} include pulsations of the AGB star which make the outflow very dynamic in the vicinity of the stars compared to the steady outflow in our simulations. 
Moreover, in their works the acceleration of the wind is driven by the pulsations of the AGB donor star and radiation pressure on dust grains, which leads to a different wind velocity profile than in Figure \ref{p2:fig:single_profile}.
These physical mechanisms may influence the amount of material that is able to reach the accretion region in their models.

In our two models in which the outflow of the gas resembles WRLOF we find that the mass accretion efficiencies are relatively low, either close to the BHL value (model Q2a5v5$\Omega$) or below it (model Q2a4v1$\Omega$; see Table \ref{p2:table:results}). 
From Figure \ref{p2:fig:beta_rot}, we observe that for $v_{\infty}/v_\mathrm{orb} < 0.3$ the results of the WRLOF simulations of  \cite{shazrene}, as reported by \cite{carlo1}, also show a much lower value than the one expected from the BHL approximation. 
This may have important consequences for the results based on this model, which is used to study the population of CEMP-$s$ stars \citep{carlo4, Abate+2018}.
From our understanding, not only the mass ratio and the orbital separation are important parameters that influence binary interaction, but as pointed out in Paper I and confirmed in the present study, the velocity of the wind plays a major role too. 
Therefore, the fit for the mass-accretion efficiency by \cite{carlo1} may only be valid for the low wind velocities studied by \citet{shazrene}, $v_\infty \approx 4$ km s$^{-1}$, rather than the 15 km s$^{-1}$ they used in their population synthesis study. 
This implies that \cite{carlo1} over-estimate the mass-accretion efficiency, which may change not only the number of CEMP stars they find, but also their final orbital-period distribution. 

\subsection{Possible implications for low-mass binary evolution}

In order to quantify the impact of our results for the mass-accretion efficiency and angular-momentum loss on the evolution of low- and intermediate-mass binary stars, the relations for these quantities given in Section \ref{p2:sec:am_nr} and Section \ref{p2:sec:beta_nr} need to be implemented in binary population synthesis codes. 
In the following we discuss how our results may impact the evolution of these binary systems.

\cite{Abate+2018} apply the fit for angular-momentum loss as a function of $v_{\infty}/v_\mathrm{orb}$ that we provide in Paper I in combination with the analytical formula of \cite{carlo1} for the WRLOF mass-accretion efficiency to a population of progenitors of CEMP stars. 
They find that this model is not able to reproduce the observed orbital periods of CEMP binary stars. 
A self-consistent model where both the angular-momentum loss and mass-accretion efficiency are derived under the same conditions may help to improve their results. 
However, we can already foresee some problems that may be encountered if we were to apply our current results.
With the canonical BHL model, population synthesis codes usually underestimate the observed frequency  of CEMP stars among very metal poor stars. 
\cite{carlo1} showed that when mass transfer occurs via WRLOF over a wide range of separations, the large associated accretion efficiencies increase the frequency of CEMP systems. 
However, as discussed in Section \ref{p2:sec:beta}, our results yield mass-accretion efficiencies that are at most a factor of $\approx 1.6$ larger than the BHL estimates and much lower than in the WRLOF model of \cite{carlo1}. 
As discussed by \citet{carlo1, Abate+2018}, the main effect of a lower accretion efficiency is that CEMP stars are produced in a narrower range of orbital periods and masses.
Consequently, the resulting CEMP fraction will be lower with our model than in the WRLOF prescription and will likely underestimate the observations. 
Another aspect that may influence the results of current binary population synthesis studies is that these typically do not include a velocity profile of the wind, but assume a constant wind velocity. 
As has been shown in Paper I and in this work, the velocity of the wind plays a major role in driving the interaction of the binary system. 

We find that the angular-momentum loss is relatively more enhanced above the isotropic-wind value for systems with lower mass ratios (i.e. more equal mass components).
With the angular-momentum loss derived from our hydrodynamical models, systems with $q=1$ and $v_{\infty}/v_\mathrm{orb} \lesssim 1.4$ will shrink instead of widening as would be predicted if the wind were spherically symmetric.
However, as $q$ increases, the maximum value of $v_{\infty}/v_\mathrm{orb}$ for which the orbit shrinks decreases. 
For instance, for $q = 4$, only systems with $v_{\infty}/v_\mathrm{orb} \lesssim 0.82$ will shrink. 
We find that 15 out of our 26 models will shrink, whereas the isotropic wind mode usually adopted in binary population synthesis codes would have predicted a widening of their orbit. 
Compared to the isotropic-wind mode, such systems will evolve towards shorter orbital periods, implying that equal-mass systems are likely to undergo a common envelope (CE) phase \citep[see e.g.][for a review on CE]{Ivanova+2013} for a wider range of initial orbital separations. 
In a test of our fits for the angular-momentum loss and mass-accretion efficiencies in the binary population synthesis code \texttt{binary\_c} we find that the maximum initial orbital separation at which systems will enter a common envelope increases by a factor between 1.25 and 1.5 AU compared to the isotropic wind case, depending on the mass ratio of the binary. 
The smallest increase from 4.38 AU to 5.46 AU, occurs for $q = 4$, and the largest shift, from 4.32 AU to 6.58 AU for $q=4/3$.
This effect may be important in determining the formation rate of the progeny of low- and intermediate-mass binary systems, such as type Ia supernovae, cataclysmic binaries and double white dwarfs.

\section{Conclusions}\label{p2:sec:conclude}

We find that the two main parameters that determine how mass transfer proceeds in low-mass binary stars interacting via AGB winds are the mass ratio of the binary system and the ratio $v_{\infty}/v_\mathrm{orb}$.
The morphology of the outflow depends on these parameters and determines the amount of mass accreted by the companion of the AGB star and the specific angular momentum that is carried away by the ejected material. 
Furthermore, because $v_\mathrm{orb}$ depends on the orbital separation, we find that modifying the orbital separation of the system produces similar results as in Paper I, where the orbital separation was fixed and the wind velocity was varied.

For any mass ratio, a low value of $v_{\infty}/v_\mathrm{orb}$ produces high mass-accretion efficiencies and large angular-momentum loss, whereas for large $v_{\infty}/v_\mathrm{orb}$ the angular-momentum loss approaches the isotropic-wind value and the mass-accretion efficiency approaches the BHL approximation.
When the masses of the stars are similar, the interaction between the wind and the companion star is stronger resulting in a larger angular-momentum loss and high accretion efficiency. 

Using \texttt{binary\_c} we model the evolution of binary systems for a large grid of orbital separations that span our hydrodynamical models. 
We find that for binary stars with initial orbital separations up to 7-10 AU, depending on the mass ratio, tidal forces are efficient and spin-orbit coupling is expected at some point during the AGB phase. 
In particular, for systems with stellar parameters similar to those in our hydrodynamical models and orbital separations smaller than  $\approx 4-6$AU the donor star is nearly in corotation with the orbit. 
Our hydrodynamical simulations show that for systems with $v_\infty/v_\mathrm{orb} \gtrsim 0.7$ rotation of the donor star has little effect on the morphology of the outflow, the mass-accretion efficiency and the orbital angular-momentum loss. 
On the other hand, for smaller velocity ratios, corotation of the AGB donor star plays an important role in the way the stars interact. 
The outflow morphologies for the two corotating models in which $v_{\infty}/v_\mathrm{orb} \lesssim 0.7$ resemble the WRLOF geometry and their orbital angular-momentum loss as well as the mass-accretion efficiencies are different from their non-rotating counterparts. 
This implies that for $v_{\infty}/v_\mathrm{orb} \lesssim 0.7$,  the orbital evolution of the system cannot be predicted by treating independently the tidal interactions and the angular-momentum lost due to material escaping the binary system, as is usually done in binary population synthesis codes.

Finally, we fit an analytical relation to our results for the angular-momentum loss and mass-accretion efficiency as a function of the mass ratio and $v_{\infty}/v_\mathrm{orb}$.
In order to test if our results help to reproduce the orbital-period distribution of the progeny of low-and-intermediate-mass binaries interacting via AGB winds, our fits for angular-momentum loss and mass-accretion efficiency need to be applied in binary population synthesis studies.

\begin{acknowledgements}
The authors thank the anonymous referee for the useful comments that improve this paper. 
MIS thanks Frank Verbunt and Richard Stancliffe for the very useful comments during the writing of this manuscript. 
MIS also thanks Glenn-Michael Oomen for the very interesting discussions during the development of this project. 
MIS is very grateful to the Stellar Dynamics and Computational Astrophysics group at Leiden, especially to Simon Portegies-Zwart and Inti Pelupessy for the the very helpful input during the progress of this project. 
This work was supported by the Netherlands School for Research in Astronomy (NOVA). 
CA is the recipient of an Alexander von Humboldt Fellowship.
\end{acknowledgements}

\bibliography{../../bibliography} 

\begin{appendix}
\section{CEMP abundances}

Table \ref{p2:table:abundances} shows the surface abundances for an AGB star with an initial metallicity of $Z = 10^{-4}$, as obtained with the method described in Section \ref{p2:sec:method}. 
These abundances are used in the implementation of the cooling of the gas. 

\begin{table}[h]
\centering
\caption{Surface abundances of the AGB donor star, for initial metallicity $10^{-4}$.}
\label{p2:table:abundances}
\begin{tabular}{c c c c}
\hline\hline		
Element  &  $\log(n/n_\mathrm{H})$ & Element  &  $\log(n/n_\mathrm{H})$  \\ \hline
He       &         -0.95      &       Al       &         -7.63 \\
C         &        -2.79       &      Si        &        -6.56 \\
N         &       -5.33        &      S          &       -6.96 \\
O         &        -4.35       &       Ar         &       -7.69  \\
Ne       &         -4.54       &      Ca        &        -7.84 \\
Na       &         -6.82       &      Fe         &       -6.59 \\
Mg       &         -6.27       &      Ni          &      -7.91 \\ \hline
\end{tabular}
\end{table}

\section{Opening angle}\label{p2:ap:mach}

\begin{figure}
\centering
\includegraphics[width=\hsize]{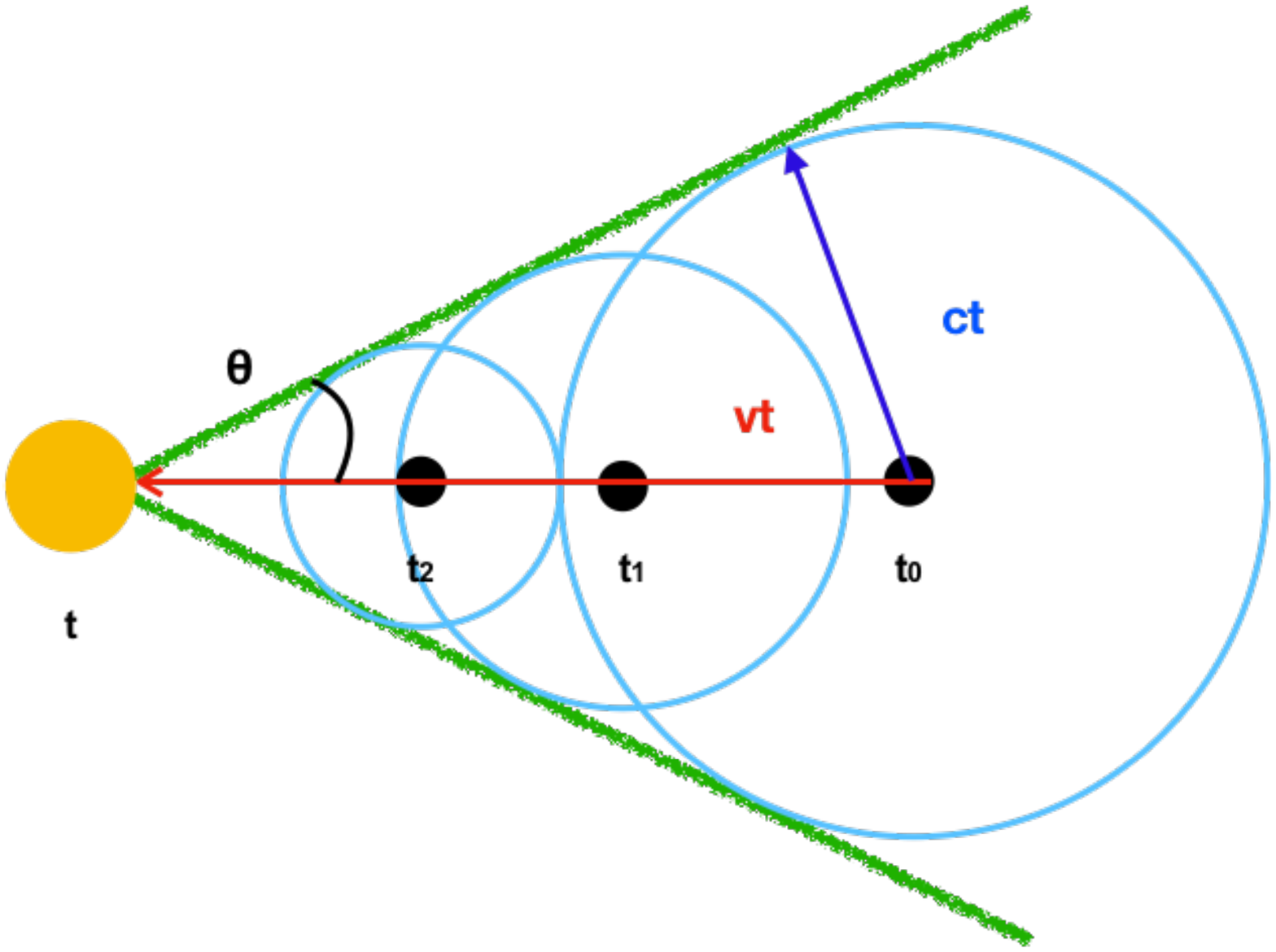}
\caption[Adapted from Prof. Jerry M. Seitzman lecture notes]{Schematic representation of the opening angle. The green lines represent the cone formed by the sound waves (light blue) propagating at different times\protect\footnotemark . } 
\label{p2:fig:opening}
\end{figure}

An object moving in a fluid causes small disturbances that propagate as sound waves with speed $c$. 
In the supersonic case $(v>c)$ the sound waves propagate within a cone delimited by tangents to the sound wave spheres. 
The angle between the line delimiting the cone and the flow moving at velocity $v$ forms what is known as the opening angle $\theta$. 
From Figure \ref{p2:fig:opening} we see that from trigonometry this angle is: $$\theta = \sin^{-1} \left(\frac{ct}{vt}\right) = \sin^{-1} \left(\frac{c}{v}\right).$$
The Mach number, $\mathcal{M}$, is defined as the ratio $v/c$. 

\footnotetext{Adapted from Prof. Jerry M. Seitzman lecture notes (\url{http://www.seitzman.gatech.edu/classes/ae2010/machanglenumber.pdf})}

\section{Dynamical change in $a$}\label{p2:ap:3}
In the Newtonian approximation, the total energy of the relative orbit of two point masses moving under their mutual gravity is given by the sum of the kinetic and potential energies. 
The energy per reduced mass is defined as:
\begin{equation}
\label{p2:eq:energy1}
\epsilon \equiv \frac{v^2}{2} - \frac{G(M_\mathrm{d} + M_\mathrm{a})}{r},
\end{equation}
where $r$ is the distance between the bodies and $v$ their relative velocity. 
The angular momentum of the binary per reduced mass is:

\begin{equation}
\label{p2:eq:ell_from_cross}
\ell = |\mathbf{r} \times \mathbf{v}|.
\end{equation}

Since the change in the orbit is small over one orbital period, we approximate the orbits as Keplerian. 
In this approximation, the relative orbit of the two masses is always a conic section.
In order to check whether the orbits of our models remain circular, we assume that the orbit of the system is an ellipse for which the orbital energy per reduced mass is:
\begin{equation}
\label{p2:eq:energy2}
\epsilon = -\frac{G(M_\mathrm{d} + M_\mathrm{a})}{2a}, 
\end{equation}
and the angular momentum per reduced mass:
\begin{equation}
\label{p2:eq:ell}
\ell^2 = \frac{[G(M_\mathrm{d} + M_\mathrm{a})]^2}{2\epsilon}(e^2-1). 
\end{equation}
At any time in our models, we estimate $e$ from Equations \ref{p2:eq:energy1}, \ref{p2:eq:ell} and \ref{p2:eq:ell_from_cross}.
$a$ can be determined at any time from the orbital energy using Equations \ref{p2:eq:energy1} and \ref{p2:eq:energy2}, or if the orbit remains circular, $a$ can be computed from the angular momentum using Equations \ref{p2:eq:ell}, \ref{p2:eq:energy2} and \ref{p2:eq:ell_from_cross}.
We estimate $\dot{a}$ as the average rate of change of $a$ over the last 5 completed orbital periods. 

\section{Spin of the AGB star}\label{p2:appendix:rot}

\begin{figure}
\centering
\includegraphics[width=\hsize]{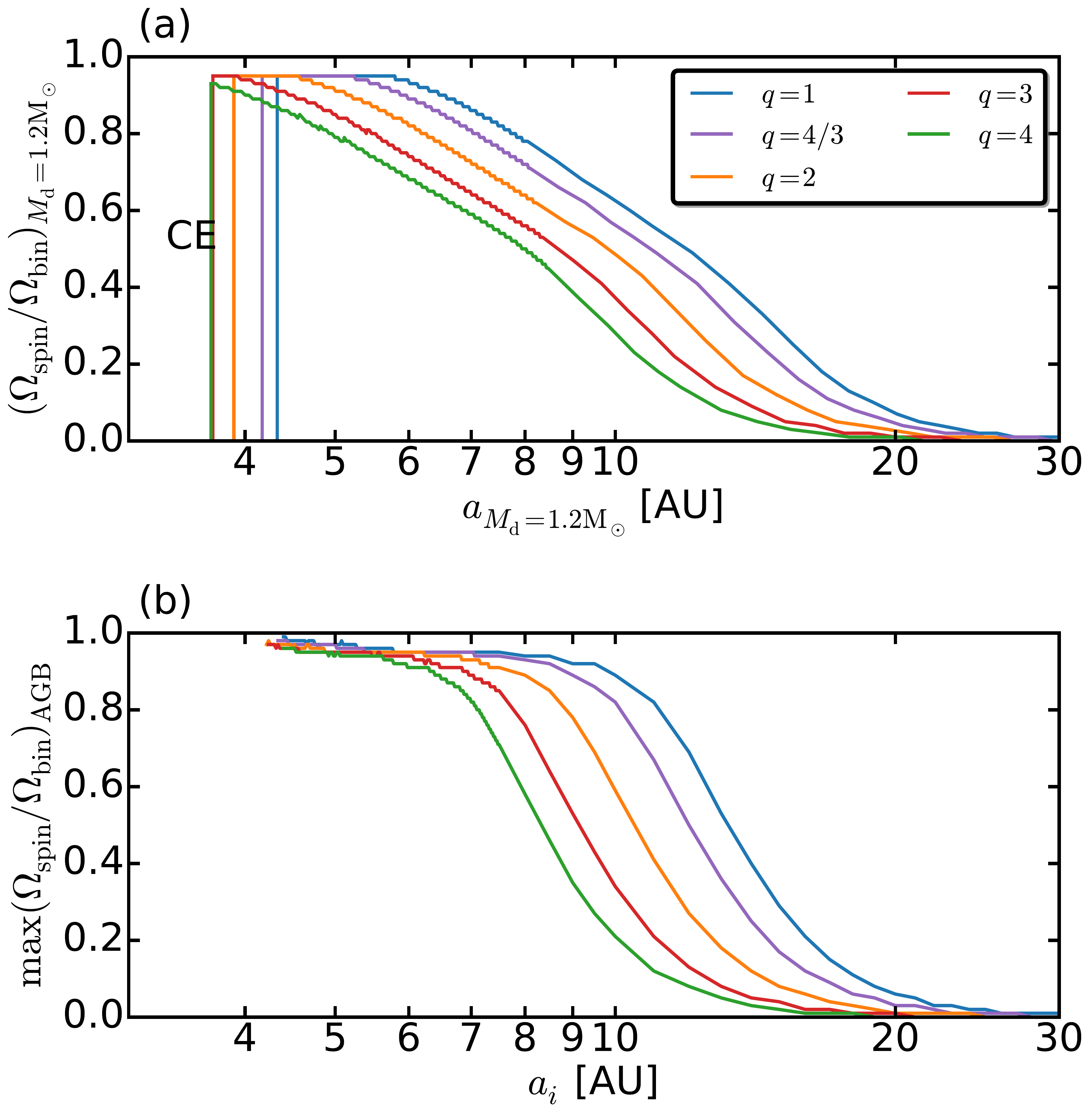}
\caption{
(a) $\Omega_\mathrm{spin}/\Omega_\mathrm{bin}$ as a function of the semi-major axis at the time the AGB star has stellar parameters as in Table \ref{p2:table:donor_star}. Colour lines represent different mass ratios. The region where the vertical lines start corresponds to orbital separations for which the binary system had gone into a common envelope (CE) at the time we computed the binary parameters. (b) Maximum $\Omega_\mathrm{spin}/\Omega_\mathrm{bin}$ as a function of the initial orbital separation of the system. Colours represent equivalent mass ratios.
}
\label{p2:fig:omegas}
\end{figure}

In this section we present the results obtained for spin-orbit coupling for the binary models evolved with \texttt{binary\_c} as described in Section \ref{p2:sec:method_rot}. 
We assume the stars to be initially non-rotating and we follow the evolution of the spin and orbital angular velocity ($\Omega_\mathrm{spin}$ and $\Omega_\mathrm{bin}$) under the influence of tidal interactions. 

Figure \ref{p2:fig:omegas}a shows the ratio $\Omega_\mathrm{spin}/\Omega_\mathrm{bin}$ as a function of the orbital separation at the moment the AGB donor star has the stellar parameters used as input for our hydrodynamical simulations (Table \ref{p2:table:donor_star}). 
The lines of different colour correspond to different mass ratios. 
We see that $\Omega_\mathrm{spin}/\Omega_\mathrm{bin}$ never reaches a value equal to one. 
However, for small orbital separations ($a \approx 4-6$, depending on mass ratio) $\Omega_\mathrm{spin}/\Omega_\mathrm{bin} \approx 0.95$, which indicates that the AGB donor is nearly in corotation with the orbit of the binary. 
We also observe that for smaller mass ratios, the tidal effects are strong at relatively larger orbital separations. 
For instance, for $a \approx 10$ AU, and $q=1$, the donor star rotates at $\approx 60\%$ of the angular velocity of the binary at the time when the mass of the donor is $M_\mathrm{d} = 1.2$ M$_{\odot}$ (Figure \ref{p2:fig:omegas}a). 
For $q=4$ at the same orbital separation, the donor star rotates only at $\approx 30\%$ of the orbital angular velocity. 
It is important to note that even for large orbital separations, tidal interaction can trigger rotation of the AGB star. 
This effect drops considerably with orbital separation and becomes negligible for $a \gtrsim 20$ AU. 

Since at the time we measure $\Omega_\mathrm{spin}/\Omega_\mathrm{bin}$, the donor star is at the superwind phase and has already gone through previous episodes of mass loss and interaction with its less evolved companion before and during the AGB phase, we also compute the maximum angular velocity that the donor star reaches during the AGB phase. 
Figure \ref{p2:fig:omegas}b shows the maximum $\Omega_\mathrm{spin}/\Omega_\mathrm{bin}$ as a function of the initial orbital separation of the binary for different mass ratios. 
We see that spin-orbit coupling can lead to near-corotation at some point during the AGB phase even for relatively wide binary stars (up to $\approx$ 10 AU), and that the orbital separation at which it occurs decreases with increasing mass ratio.
By the time the superwind stage is reached (Fig \ref{p2:fig:omegas}a) the star has spun down as a result of strong mass loss and corotation is lost again, except in the closest orbits. 

Finally, we also explore the case in which the primary star is initially rotating. 
We arbitrarily set the initial rotational velocity of the donor star to be 200 km s$^{-1}$. 
We find that at the moment the donor star has the stellar parameters used as initial set up of our SPH simulations (see Table \ref{p2:table:donor_star}), the angular velocity of the AGB star has a similar trend as shown in Figure \ref{p2:fig:omegas}a as a function of the orbital separation for a given mass ratio, up to about 20 AU.
However, for orbital separations larger than 20 AU the spin of the AGB star does not tend to zero, but the star is rotating slowly at about $\approx 0.1 \Omega_\mathrm{bin}$. 
This shows that for $a \lesssim 20$ AU the spin of the star is dominated during the AGB phase by tidal evolution and is independent of the initial spin of the donor star.

\end{appendix}

\end{document}